\documentclass[%
 reprint,
superscriptaddress,
 amsmath,amssymb,
 aps,
]{revtex4-1}

\usepackage{subfigure}
\usepackage{graphics}     
\usepackage{amsmath}
\usepackage{graphicx}
\usepackage{dcolumn}
\usepackage{bm}
\usepackage{hyperref}
\hypersetup{colorlinks=true, citecolor=blue, urlcolor=blue, linkcolor=blue}
\usepackage{array}
\usepackage{booktabs}
\usepackage[mathlines]{lineno}

\begin{document}
\preprint{APS/123-QED}

\title{Multi-stability in Doubochinski's Pendulum}

\author{Yao Luo}
\affiliation{School of Physics, Nanjing University, Nanjing, P.R China, 210093}
\author{Wenkai Fan}%
\affiliation{%
Kuang Yaming Honors School, Nanjing University, Nanjing, P.R China, 210093}%
\affiliation{%
Department of Physics, Duke University}%

\author{Chenghao Feng}%
\affiliation{%
Kuang Yaming Honors School, Nanjing University, Nanjing, P.R China, 210093}%
\affiliation{%
Department of Electrical and Computer Engineering, The University of Texas at Austin.}%

\author{Sihui Wang}\email{wangsihui@nju.edu.cn}
\affiliation{School of Physics, Nanjing University, Nanjing, P.R China, 210093}

\author{Yinlong Wang}%
\affiliation{School of Physics, Nanjing University, Nanjing, P.R China, 210093}

\date{\today}

\begin{abstract}
The widespread phenomena of multistability is a problem involving rich dynamics to be explored.  In this paper, we study the multistability of a generalized nonlinear forcing oscillator excited by $f(x)cos \omega t$. We take Doubochinski's Pendulum as an example. The so-called "amplitude quantization", i.e., the multiple discrete periodical solutions, is identified as self-adaptive subharmonic resonance in response to nonlinear feeding. The subharmonic resonance frequency is found related to the symmetry of the driving force: odd subharmonic resonance occurs under even symmetric driving force and vice versa. We solve the multiple periodical solutions and investigate the transition and competition between these multi-stable modes via frequency response curves and Poincare maps. We find irreversible transition between the multistable modes and propose a multistability control strategy.

\end{abstract}

\pacs{Valid PACS appear here}
\maketitle

\section{\label{sec:level1}Introduction}

The phenomenon of multistability (coexistence of different attractors for a given set of parameters) is widespread in physics \cite{maurer1980effect,brun1985observation,gibbs2012optical}, chemistry \cite{aguda1987bistability,wilhelm2009smallest}, biology \cite{angeli2004detection,ozbudak2004multistability,freyer2011biophysical} and climate systems \cite{robinson2012multistability,freire2008multistability} and many other fields of science and nature.  One important feature of such systems of high order of complexity are their extreme sensitivity to perturbation and initial conditions\cite{pisarchik2014control}.
Generalized multistability, firstly distinguished from ordinary bistability extensively exist in nonlinear systems, was found in laser physics\cite{arecchi1982experimental}.

The mechanisms of multistability include coupling \cite{maistrenko2007multistability}, delayed feedback \cite{balanov2005delayed,foss1996multistability},parametrical forcing \cite{varma1993quadratic} and dissipation\cite{lieberman1985transient}. These mechanisms usually do not act independently in a system. In many cases, the feedback and dissipation often coexist. For example, a mechanical rotor with weak damping and self-adaptive driving term possess over one hundred attractors\cite{feudel1996map}.

Compared to bistability phenomenon, the study of generalized multistability is far from being complete. The phenomena and mechanisms of multistability involve rich dynamics to be explored. Here we study the multiple stability behaviors in a classical mechanical system, the Doubochinski's pendulum\cite{tennenbaum2006amplitude}. We reveal that it's mysterious "amplitude quantization", initially studied as "macroscopic quantum behavior"\cite{tennenbaum2006amplitude} is, in fact, the self-adaptive behavior in response to nonlinear feeding.
　
 We consider a generalized model of an oscillator excited by periodic force $f(x)cos \omega t$ where the feeding function $f(x)$ is a nonlinear function of displacement. This model containing restoring force, dissipation and nonlinear feeding can be regarded as generalization of parametric oscillation. For a linear parametric oscillator governed by Mathieu equation, subharmonic resonance occurs when the driving frequency is near twice the natural frequency of an oscillator \cite{nayfeh2008nonlinear}.
To understand the mechanism of multistability, we express the feeding function $f(x)$ as series of polynomial and solved the amplitudes and frequency respond curves analytically. The origin of multiple solutions and strong self-adaptivity can be explained.  Moreover, the even-odd correspondence between the symmetry of the driving force and subharmonic resonance frequencies are presented numerically.
To investigate the transition and competition between the multi-stable modes, we present the frequency response curves and Poincare maps near $\omega =3\omega_0$ and $\omega =5\omega_0$. The transition between multistable states is more complex and intriguing than in bistability.  We find irreversible transition between the multistable modes and propose a multistability control strategy. In contrast, the frequency response for bistability states forms a closed hysteresis loop rather than an open route.

　　
\section{\label{sec:level2}Theoretical Model}
\subsection{Dynamic Equation}
A general form of nonlinearly excited oscillator can be described by the following dynamic equation
\begin{equation}\label{differential_equation_1}
\ddot{x}+2h\dot{x}+f_{r}(x)=f(x)cos \omega t
\end{equation}
where $x$ is the displacement, $h$ is the linear damping coefficient, $f_{r}(x)$ is the restoring force of the system, $f(x)cos\omega t$ is a nonlinear periodic driving force and $\omega$ is the driving frequency. $f(x)$ is the displacement-dependent amplitude of the driving force, called feeding function here. There are various physical systems described by equations of similar forms. For example, when $f_{r}(x)$ contains a positive linear term and a cubic term, and the feeding function $f(x)$  is a constant, Eq.\ref{differential_equation_1} becomes Duffing's equation \cite{kovacic2011duffing}. A nonlinear parametric oscillator is described by an equation similar to Eq.\ref{differential_equation_1}, where $f(x)$ is proportional to displacement. \cite{rhoads2006generalized}.

When applied to a pendulum the restoring force can be written as
\begin{equation}\label{differential_equation_2}
f_r(x)= \omega_{0}^{2}sin(x)
\end{equation}
where $\omega_{0}$ is the pendulum's natural frequency and $x$ is the angular displacement.

The actual form of the feeding function is cumbersome, so we approximate the feeding function in terms of polynomial of $x$ to do analytical calculation. Specifically, if $f(x)$ is an even symmetric function of $x$, the polynomial approximation only contains even-degree terms,
\begin{equation}\label{eq3}
f_{even} (x)= a_0 +a_2 x^2 + a_4 x^4+...
\end{equation}
If $f(x)$ is an odd symmetric function of $x$, it only contains odd-degree terms.
\begin{equation}\label{eq4}
f_{odd}(x) = a_1 x + a_3 x^3+...
\end{equation}

In many cases, the feeding function $f(x)$ is considerable in an active zone and becomes negligible elsewhere, such a pendulum is called a "kick-excited" one. In Ref.  \cite{damgov2000discrete}, a symmetric $\Pi$ shaped feeding function is considered. Here, the polynomial series is truncated after the lowest few terms to model the feeding function. The coefficients are determined by fitting $f(x)$ of a particular physical model using above polynomials. Since the polynomial usually does not converge to the practical feeding function for arbitrary displacement, the polynomial approximation only applies to the range of oscillation or the "active zone" of the feeding function unnecessarily being small.

\begin{figure}
  \centering
   \subfigure[]{
    \label{1-0} 
    \includegraphics[width=0.2\textwidth]{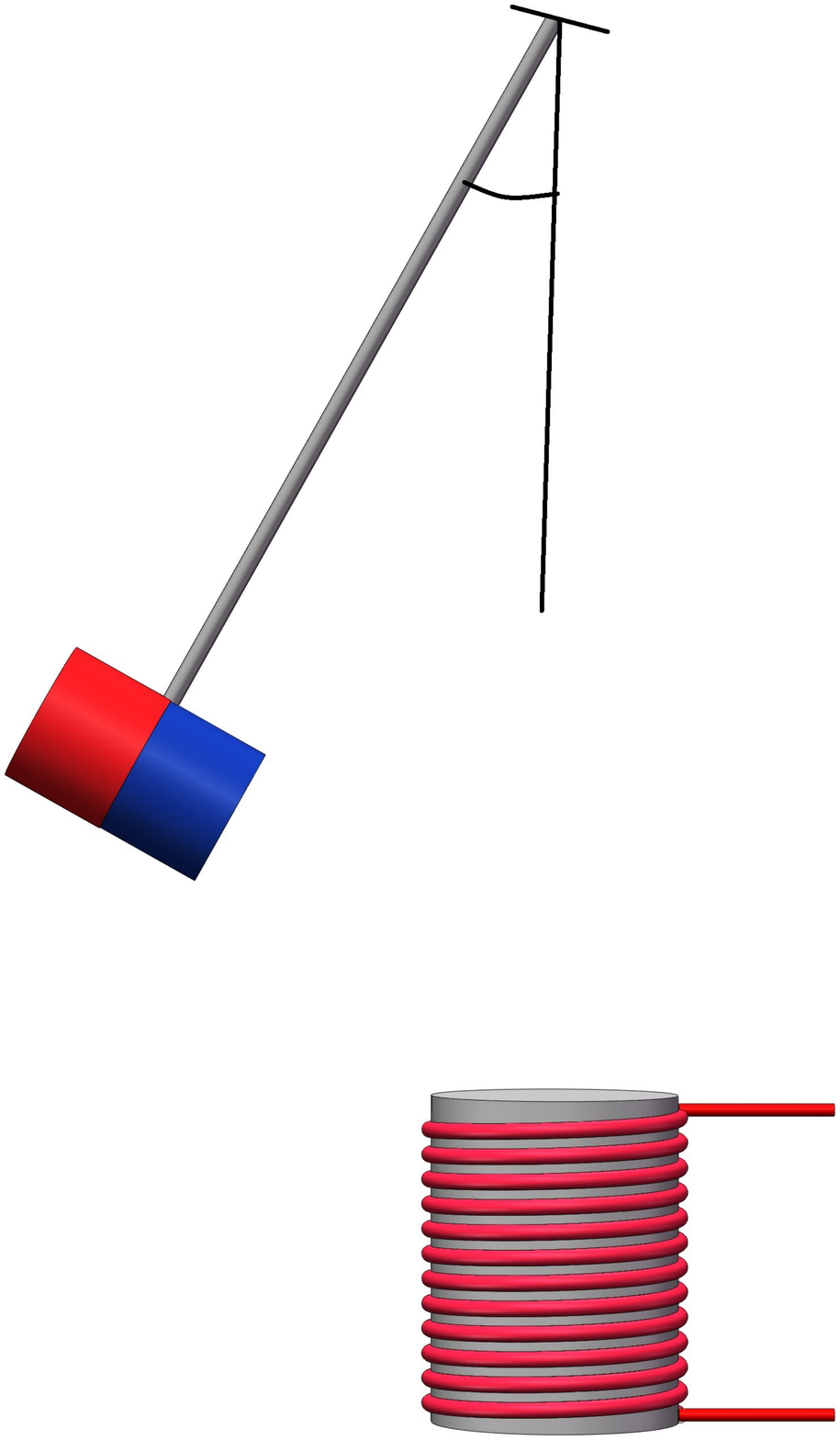}}
   \subfigure[]{
    \label{1-1} 
    \includegraphics[width=0.2\textwidth]{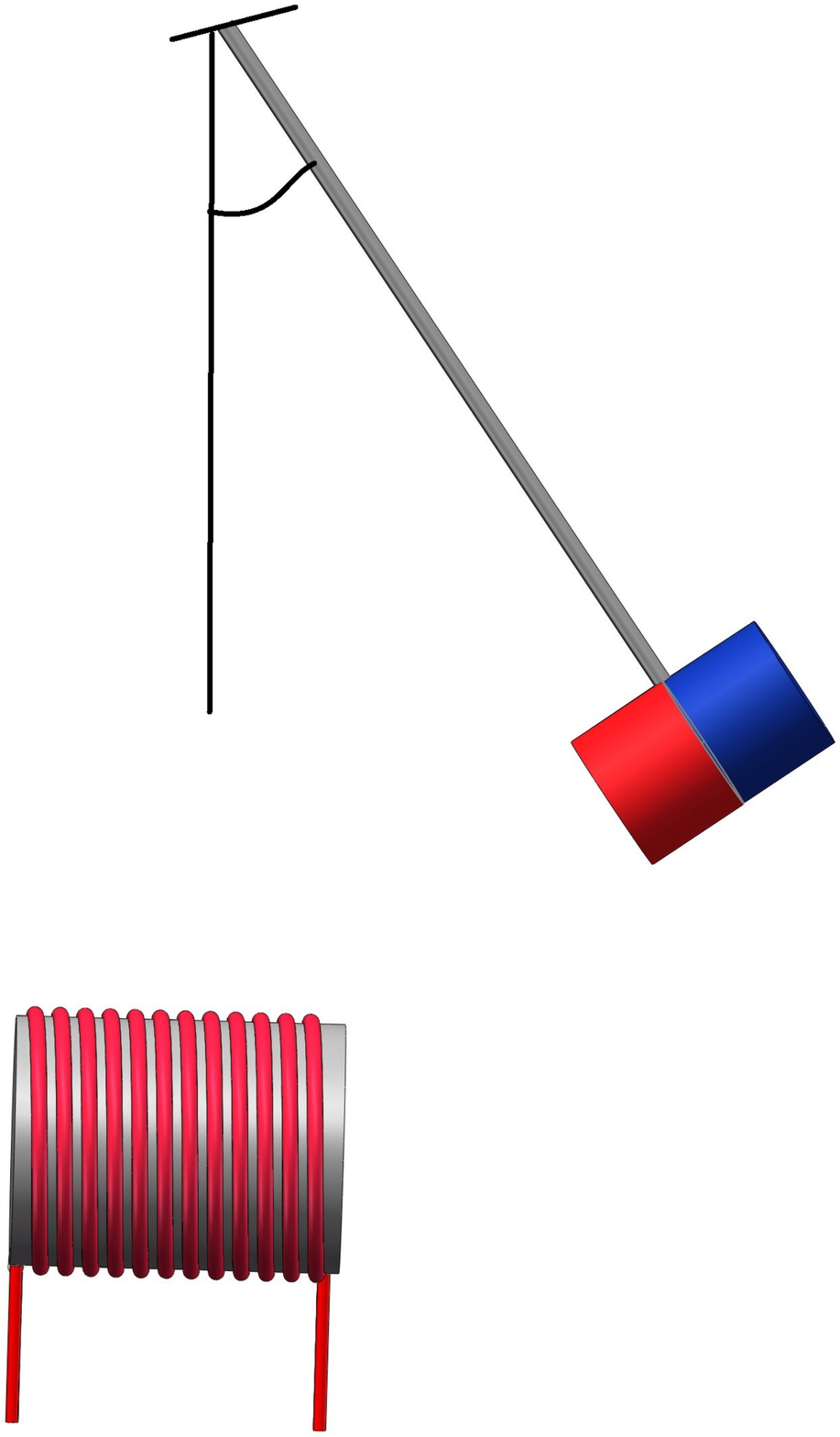}}
   \subfigure[]{
    \label{1-2} 
    \includegraphics[width=0.2\textwidth]{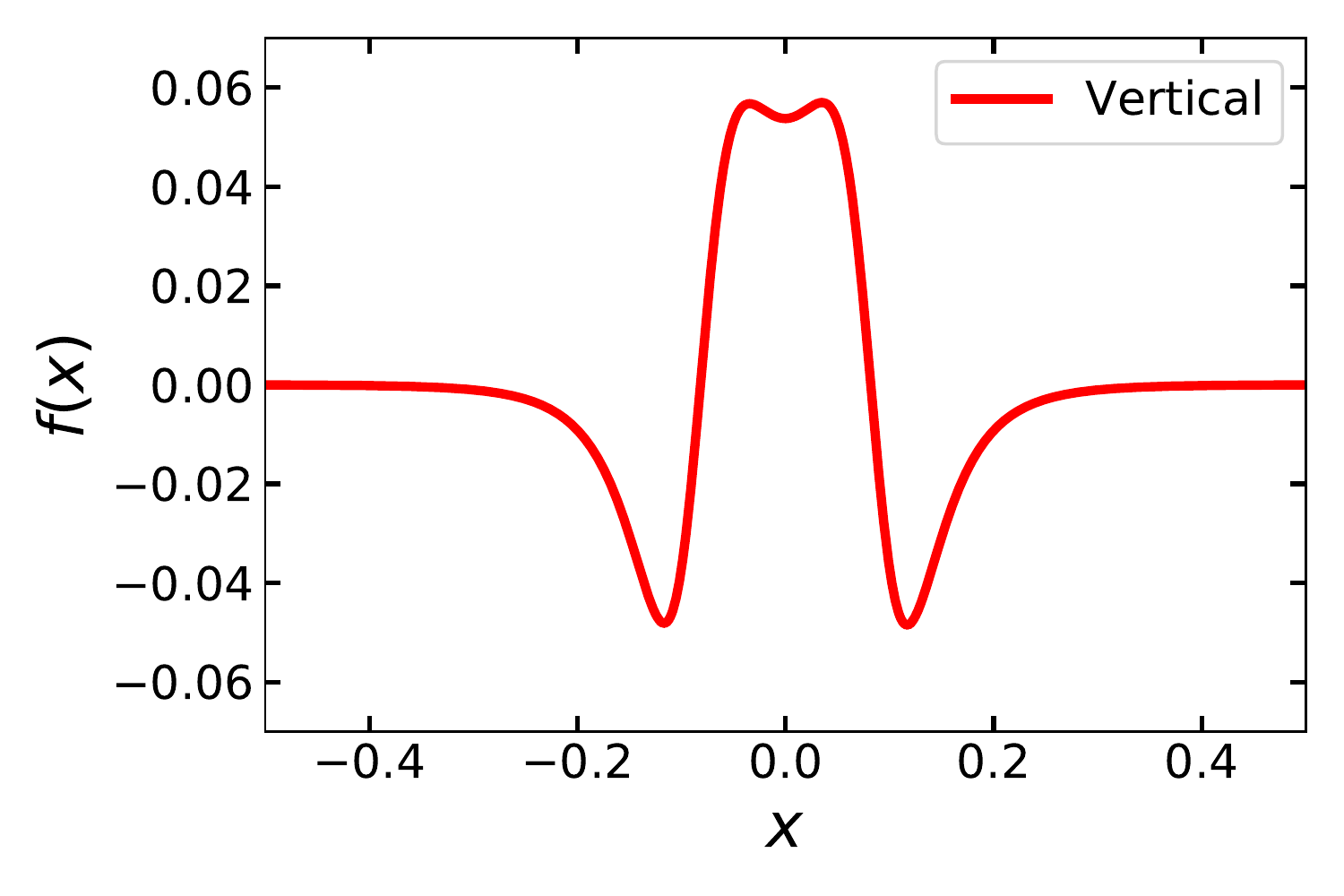}}
  \subfigure[]{
    \label{1-3} 
    \includegraphics[width=0.2\textwidth]{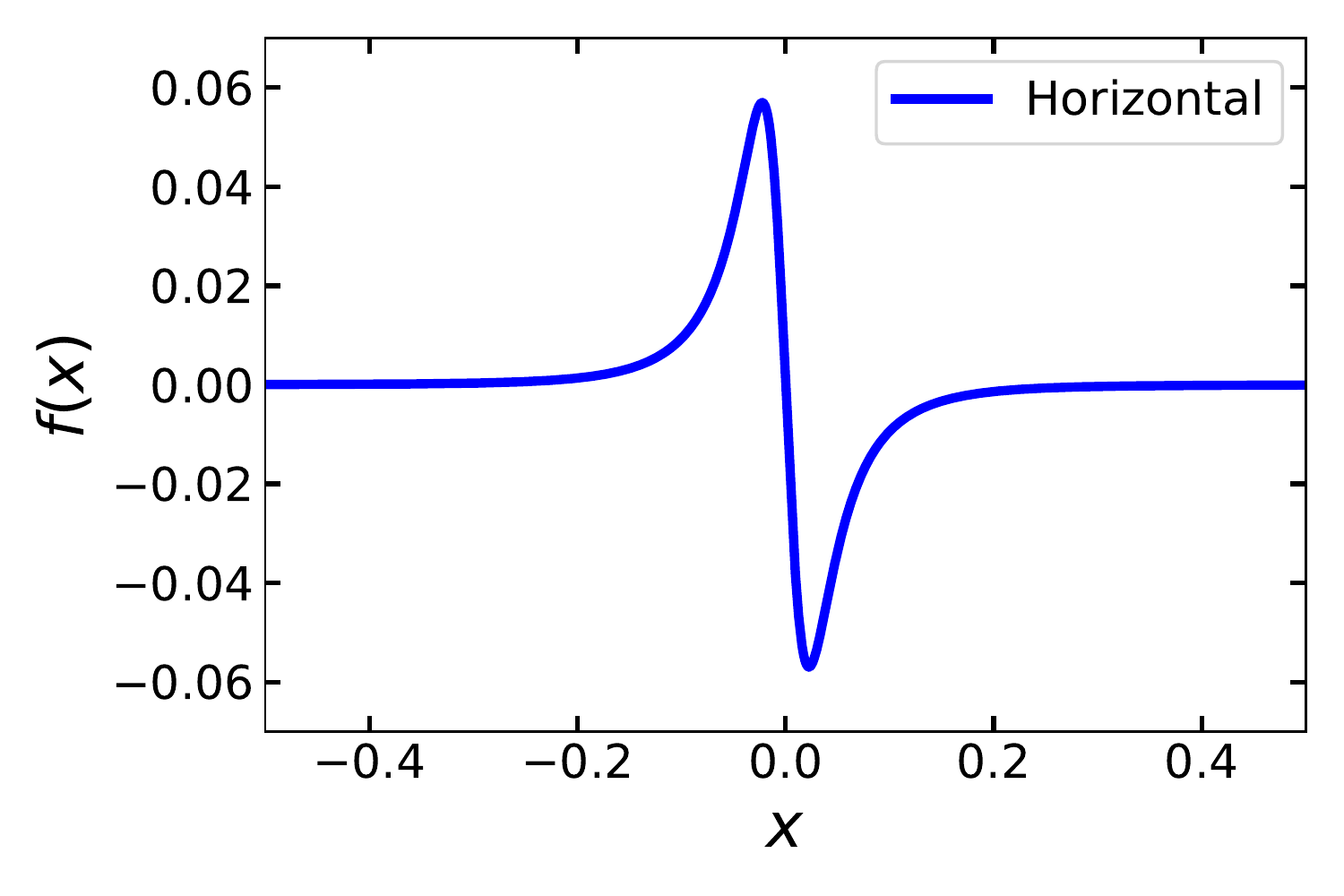}}

  \caption{\textbf{a},\textbf{b}, Schematic diagrams of the Doubochinski’s pendulum.
The electromagnet is vertically(a) or horizontally(b) placed beneath the pendulum.
\textbf{c},\textbf{d}, Feeding function of Doubochinski’s pendulum for vertically(c) or horizontally(d) placed electromagnet.
  }

  \label{Fig_1}
\end{figure}

\subsection{Feeding function for a magnetic pendulum}
Later on, we consider a practical model of a nonlinearly excited oscillator, the Doubochinski's pendulum. The pendulum consists of a light rigid pendulum with a small magnet at the free end. An AC powered electromagnet is vertically or horizontally placed beneath the pendulum, see Figs.\ref{1-0} and \ref{1-1}. The magnet pole's direction should be adjusted along the local geomagnetic field to eliminate the torque. The dynamic equation for the pendulum is

\begin{equation}\label{eq5}
   \ddot{x}+2 h \dot{x}+\omega_{0}^{2}sinx=\frac{T(x)}{I}cos\omega t
\end{equation}
where $I$ is the moment of inertia of the pendulum, $T(x)$ is the torque applied to the magnet. We can write the dynamical equation in a dimensionless form.
\begin{equation}\label{eq6}
\frac{d^2x}{d\tau^2}+2\beta\frac{dx}{d\tau}+sinx=f(x)cos\Omega \tau
\end{equation}

where $\tau$ , $\beta$ and $f(x)$ are defined as
\begin{eqnarray}\label{eq_7Parameter}
  \tau = \omega_0 t \qquad  \beta = \frac{h}{\omega_0}\\
  f(x) = \frac{T(x)}{I\omega_0 ^2} \qquad \Omega = \frac{\omega}{\omega_0}
\end{eqnarray}

The driving term arises from the alternating magnetic field. The torque can be evaluated by treating the magnet as a magnetic dipole, thus,
\begin{equation}\label{eq_8Driving force}
   T(x)=\frac{\partial(M \cdot B)}{\partial x}
\end{equation}

where $B$ is the magnetic induction of the field, and $M$ is the magnetic moment of the magnet.
Here we calculate the feeding function in two typical configurations, parameters for calculation are given in table \ref{table}, involving natural frequency of the pendulum $\omega_0$, moment of inertia $I$, length of the pendulum $L$, radius of the coils $r$, number of turns $N$,  height of the electromagnet coil $H$, distance between the coil and the magnet $d$, magnetic momentum of the magnet $M$ and the current $i$. These physical parameters will be used for numerical calculation.
The coefficients of the polynomial $a_0$, $a_2$, $a_4$ in Eq.\ref{eq3} fitting to the feeding function given by above parameters are also listed, which will be used in analytical calculation. As shown in Figs.\ref{1-2} and \ref{1-3}, when the electromagnetic coil is placed vertically, $f(x)$ has even symmetry; when the coil is placed horizontally, $f(x)$ has odd symmetry.

\renewcommand\arraystretch{1.3}
\begin{table}[h]
  \centering
    \caption{The parameters used to calculate the feeding function
and the fitting coefficients of the polynomial.}\label{table}
  \squeezetable
  \begin{ruledtabular}
\begin{tabular}{lcccc} 

\multicolumn{5}{l}{\textbf{Parameters : }}\\
\hline
$\omega_0(rad/s)$&$I(kg \cdot m^2) $  & $L (m)$ & $r(m)$ & $N$   \\
\hline
5.13      &0.01          &0.456    & 0.042    &220   \\
\hline
$H(m)$ &$d(m)$ & $M(A \cdot m^2)$ &$i(A)$ & $\beta(rad/s)$   \\
\hline
 0.02 & 0.021  & 1.36    & 0.5 & 0.0156 \\
\hline
\multicolumn{5}{l}{\textbf{fitting coefficients of the polynomial : }}\\
\hline
$a_0$&$a_2$&$a_4$& &  \\
\hline
0.055&-11.6&300& & \\
\end{tabular}

\end{ruledtabular}
\end{table}


\section{Dynamic Behaviors}\label{Sec3}
\subsection{Discrete Periodic Orbits}\label{Discrete}



The pendulum's motion is governed by Eq.\ref{eq6} and Eq.\ref{eq_8Driving force}. We solve the equations when the electromagnet is vertically placed, using the feeding function as shown in Fig.\ref{1-2} and Runge-Kutta 4th order method in c++.

\begin{figure}
  \centering
   \subfigure[$\Omega=2.98$]{
    \label{3-1} 
    \includegraphics[width=0.2\textwidth]{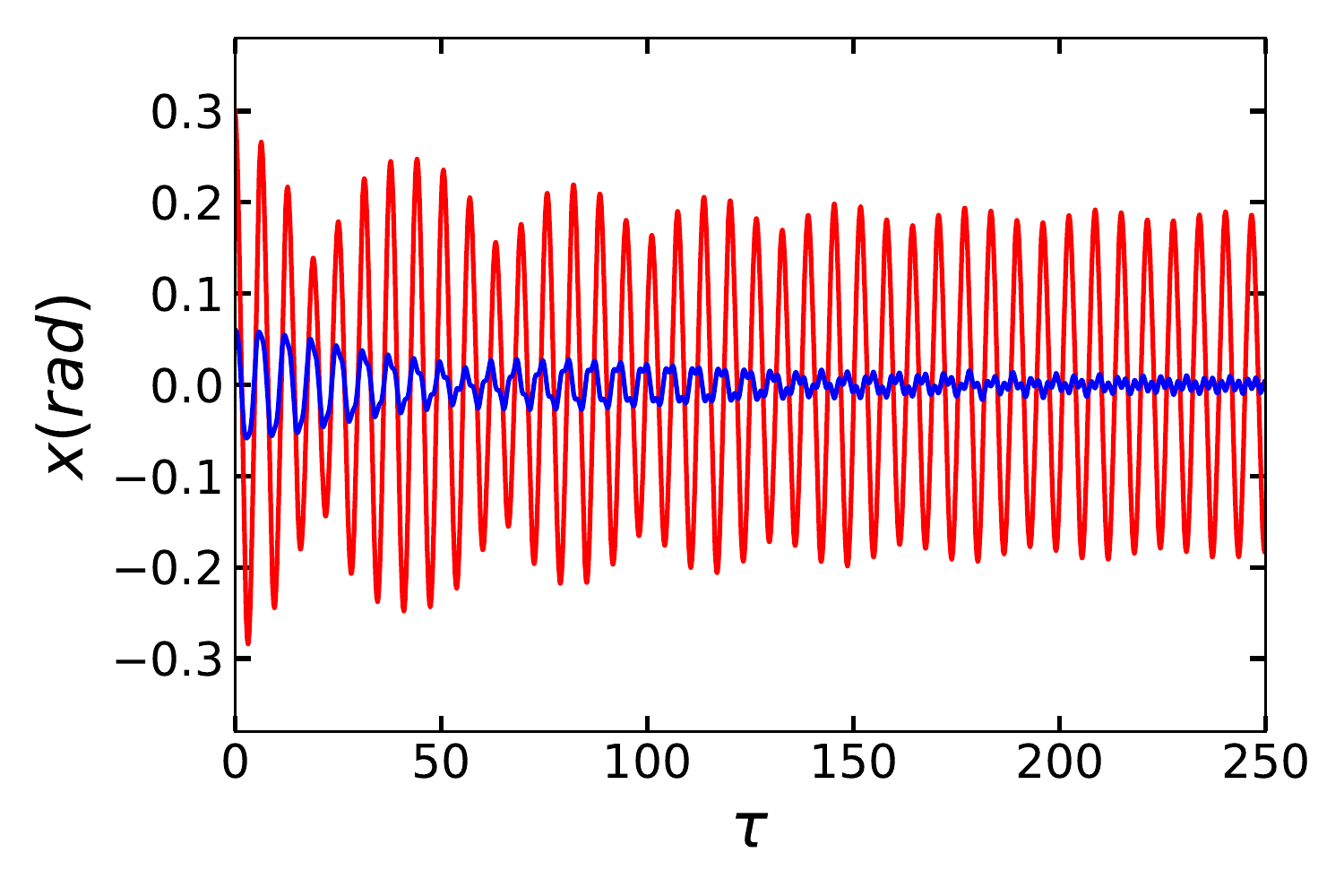}}
  \subfigure[$\Omega=4.98$]{
    \label{3-2} 
    \includegraphics[width=0.2\textwidth]{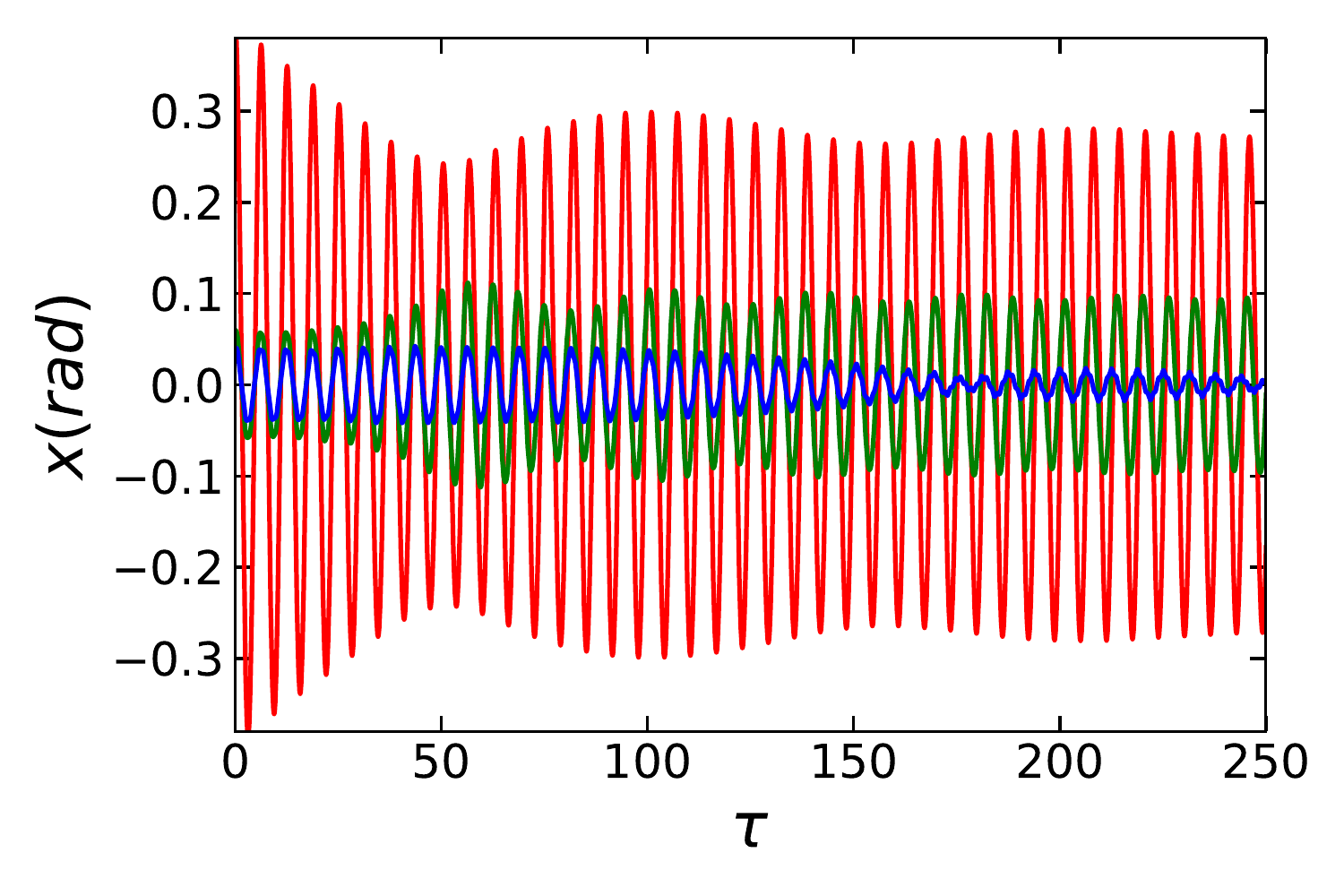}}
   \subfigure[$\Omega=2.98$]{
    \label{4-1} 
    \includegraphics[width=0.2\textwidth]{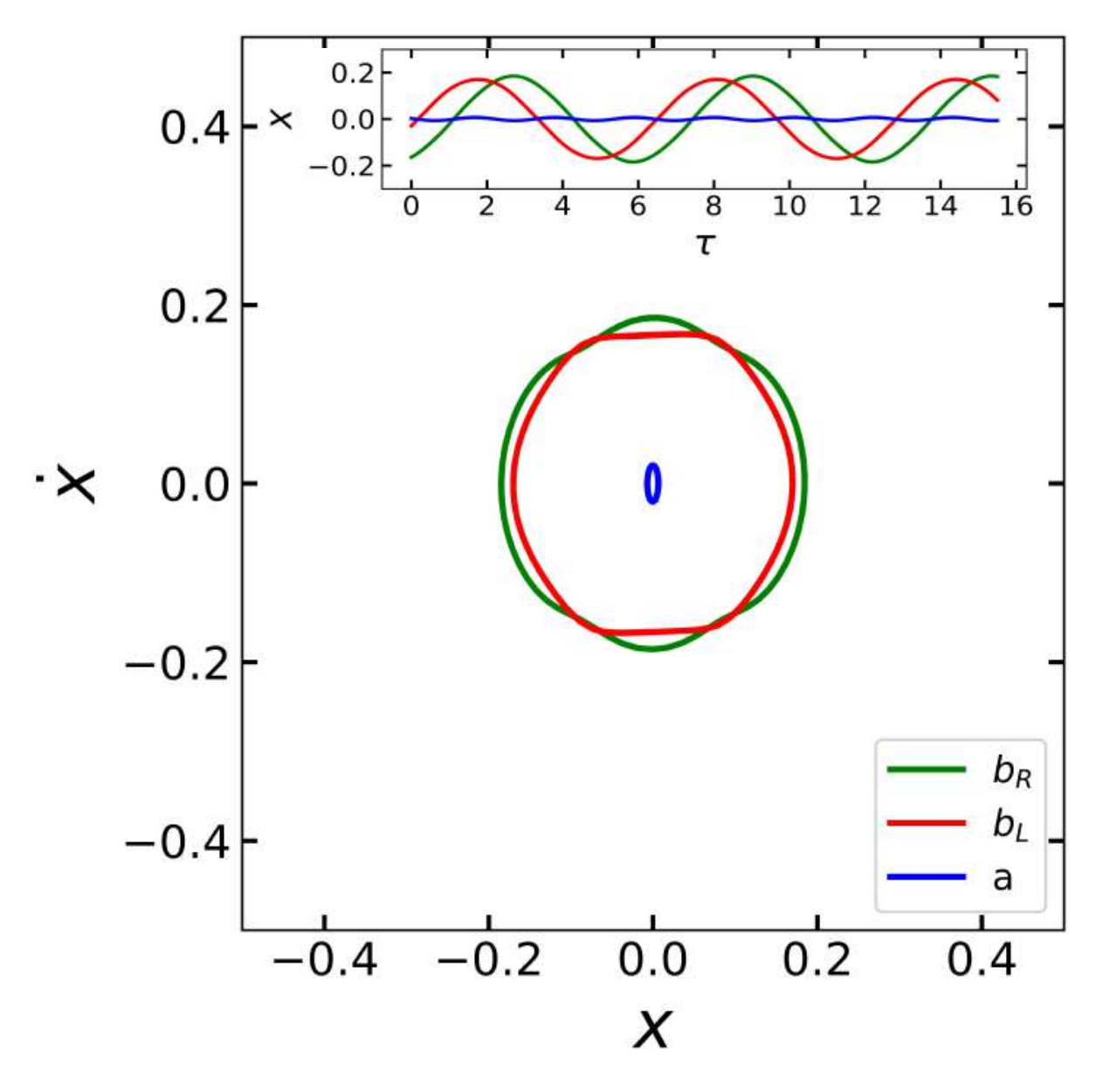}}
  \subfigure[$\Omega=4.98$]{
    \label{4-2} 
    \includegraphics[width=0.2\textwidth]{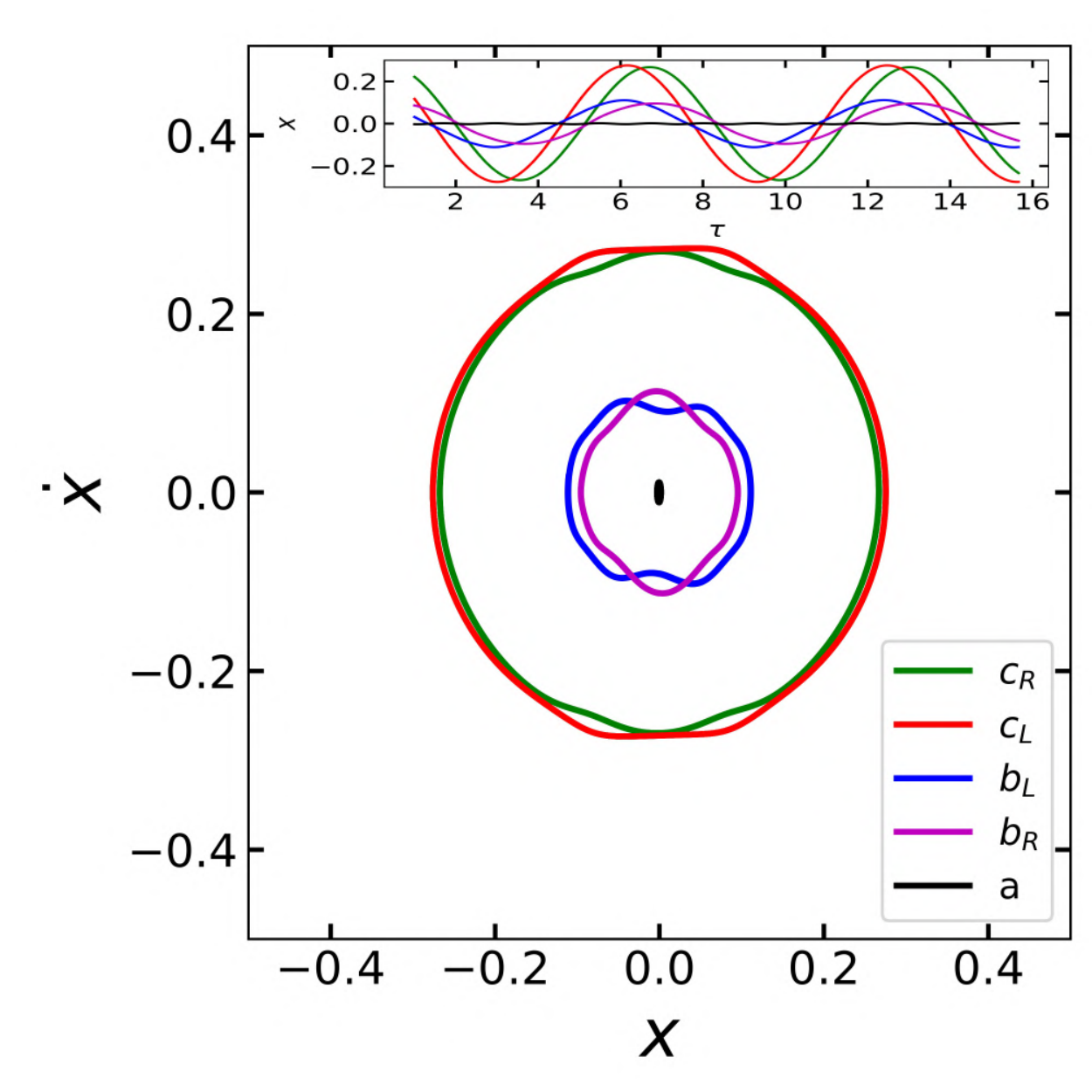}}

 \caption{\textbf{a},\textbf{b}, Multiple stationary solutions of Doubochinski's pendulum for driving frequency $\Omega=2.98, 4.98$.  The final oscillation modes vary with initial conditions. \textbf{c},\textbf{d}, Limit cycles for $\Omega=2.98, 4.98$ respectively. The small cycles in the middle are the linear solutions. The larger cycles which appear in pairs are nonlinear solutions where the pendulum oscillates near the natural frequency. The insets show stationary solutions of the corresponding limit cycles in the same color. }
  \label{Fig_3}
\end{figure}

In Fig.\ref{Fig_3}, we find that there are several stationary oscillation modes near triple ($\Omega=2.98$) and quintuple of the natural frequency ($\Omega=4.98$).Figs.\ref{3-1} and \ref{3-2} are the time evolutions of typical periodic orbits.  The final oscillation modes are dependent on initial conditions. Then we sample the initial condition in phase space of $(x, \dot{x})$, in search of stable oscillation orbits. All the trajectories converge to several limit cycles, as shown in Figs.\ref{4-1} and \ref{4-2}.
The small cycles in the middle are the linear solutions where the pendulum oscillates at the frequencies of the driving force. The larger cycles are nonlinear solutions where the pendulum oscillates near the natural frequency. The insets of Figs.\ref{4-1} and \ref{4-2} show the dual stationary solutions corresponding to the larger limit cycles. Though the two limit cycles are adjacent in phase diagram, they are different in phase. From Fig.\ref{Fig_3}, we also see that for the large stationary solutions, the pendulum approximately performs harmonic oscillation.

These results are consistent with experimental phenomenon \cite{njubook} that pendulum released at small angles only does small oscillations in the "linear region"; and for larger releasing angles, the amplitude varies slowly and ends up with stable harmonic oscillations on the large amplitude orbits or the small linear orbit.

To find the multiple periodic solutions analytically, we use a polynomial approximation of the feeding function Eq.\ref{eq3} and truncate it after the 4th order.

In Eq.\ref{eq3}, when $x<<1$, the leading term is a constant $a_0$. The pendulum only does regular linear forced oscillation at the driving frequency. At larger angles, the effect of nonlinearity becomes significant. When the pendulum performs large amplitude oscillation under driving frequency $\Omega \approx 3$, the zeroth order solution of Eq.\ref{eq6} should be written as
\begin{equation}\label{eq11}
x_0 = Acos(\frac{\Omega}{3}\tau+\phi)
\end{equation}

In order to capture the frequency response of the system, a factor $\sigma$ is added to the triple natural frequency

\begin{equation}\label{eq12}
\Omega =3 + \sigma
\end{equation}

Substituting Eq.\ref{eq11} into Eq.\ref{eq6} and averaging the amplitude and phase change over one period, we get the time derivatives of amplitude and phase,

\begin{equation}\label{eq13}
\dot{A}=-\frac{2A\beta}{1+\sigma}-\frac{A^2(4a_2+3A^2a_4)sin(3\phi)}{16(1+\sigma)}
\end{equation}
\begin{equation}\label{eq14}
\dot{\phi} = -\frac{A^2+16\sigma}{8(1+\sigma)}-\frac{A(4a_2+5A^2a_4)cos(3\phi)}{16(1+\sigma)}
\end{equation}

For simplicity, we set the damping coefficient $\beta=0$. These two equations approximately describe the evolution trajectories in $A-\phi$ space. The stationary solutions are the fixed points in $A-\phi$ space, which satisfy

\begin{equation}\label{eq15}
\dot{A}=0, \qquad \dot{\phi}=0
\end{equation}
The solutions are

\begin{center}
\begin{eqnarray}
  4a_2A+5a_4A^3=32\sigma+2A^2 , \ \phi = 2k \frac{\pi}{3}  \label{eq16}  \\
  4a_2A+5a_4A^3=-32\sigma-2A^2 ,     \  \phi = (2k+1) \frac{\pi}{3}  \label{eq17}
\end{eqnarray}
\end{center}
in which $k \in \mathbb{Z}$.



\begin{figure}
  \centering
  \includegraphics[width=0.3\textwidth]{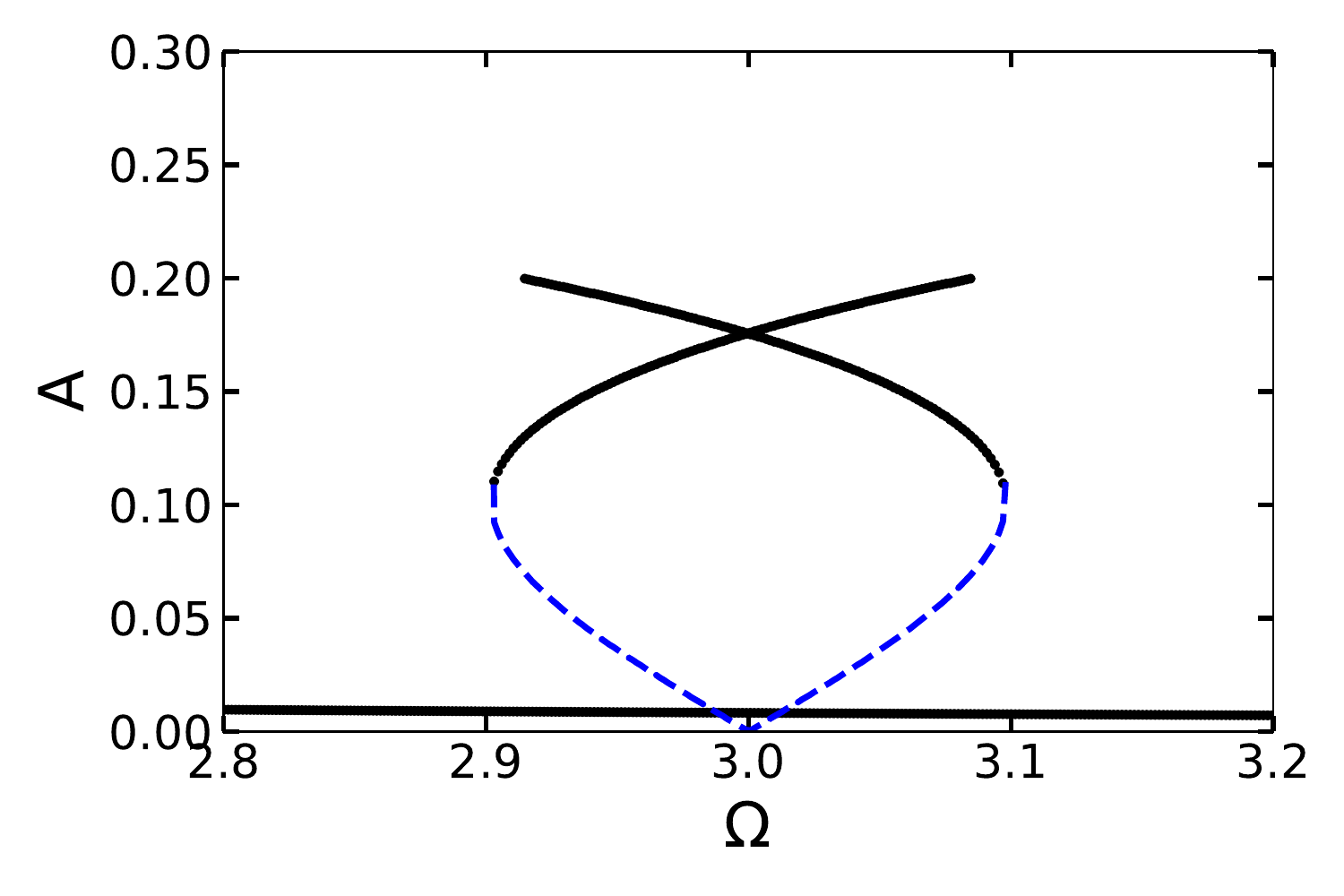}
\caption{The analytical result of amplitude frequency response curves for $\Omega \approx 3$. The two black solid curves denote the larger stable roots, the blue dashed the unstable root. The linear forced oscillation solution is the solid line near the bottom.}\label{Fig_5}
\end{figure}

The term $2A^2$ can be ignored in further analysis because it is much smaller than $a_2 A$ and $a_4 A^3$, see table.\ref{table}.
When $\sigma =0$, the amplitudes solved from Eq\ref{eq16} and \ref{eq17} are the same,
\begin{equation}\label{eq18self-adaptive}
  A =\pm \sqrt{\frac{-4a_2}{5a_4}} \ ,  0
\end{equation}

Eq.\ref{eq18self-adaptive} indicates that nontrivial stationary solutions exist when $a_2$ and $a_4$ are different in sign. Here multi-stability is achieved without dissipation when the polynomial of the feeding function contains more than one nonlinear terms so that energy balance can be realized. Feeding function truncated after the 4th order can produce resonance at $\Omega \approx 3$. For higher subharmonic frequencies, feeding function with higher order terms should be taken into consideration.

The oscillation amplitudes in Eq.\ref{eq18self-adaptive} are independent of the amplitude of the driving force, since the $a_2$,$a_4$ are proportional to the amplitude of the driving force and the ratio will not change. Such strong self-adaptivity is verified in Ref.\cite{damgov2000discrete,njubook}, in which the amplitudes almost do not change with the driving force within a very large range. Actually, as amplitude of the driving force exceeds a critical value, the oscillator will undergo a series of symmetry breaking oscillation modes with multiple period and finally becomes chaotic. The period-3 bifurcations in a system with  $\Pi$ shaped feeding function reported by Domgov \cite{damgov2000discrete} is a similar example.

The frequency response curves shown in Fig.\ref{Fig_5} can be solved from Eq.\ref {eq17} when $\sigma \neq 0$. The nonlinear solutions contain one smaller unstable root, denote by blue dotted curve, and a pair of larger stable roots, denote by black solid curves. The linear forced oscillation solution is also plotted, see the solid line near the bottom in Fig.\ref{Fig_5}.



\begin{figure}[htpb]
  \centering
   \label{6-1}
    \includegraphics[width=0.4\textwidth]{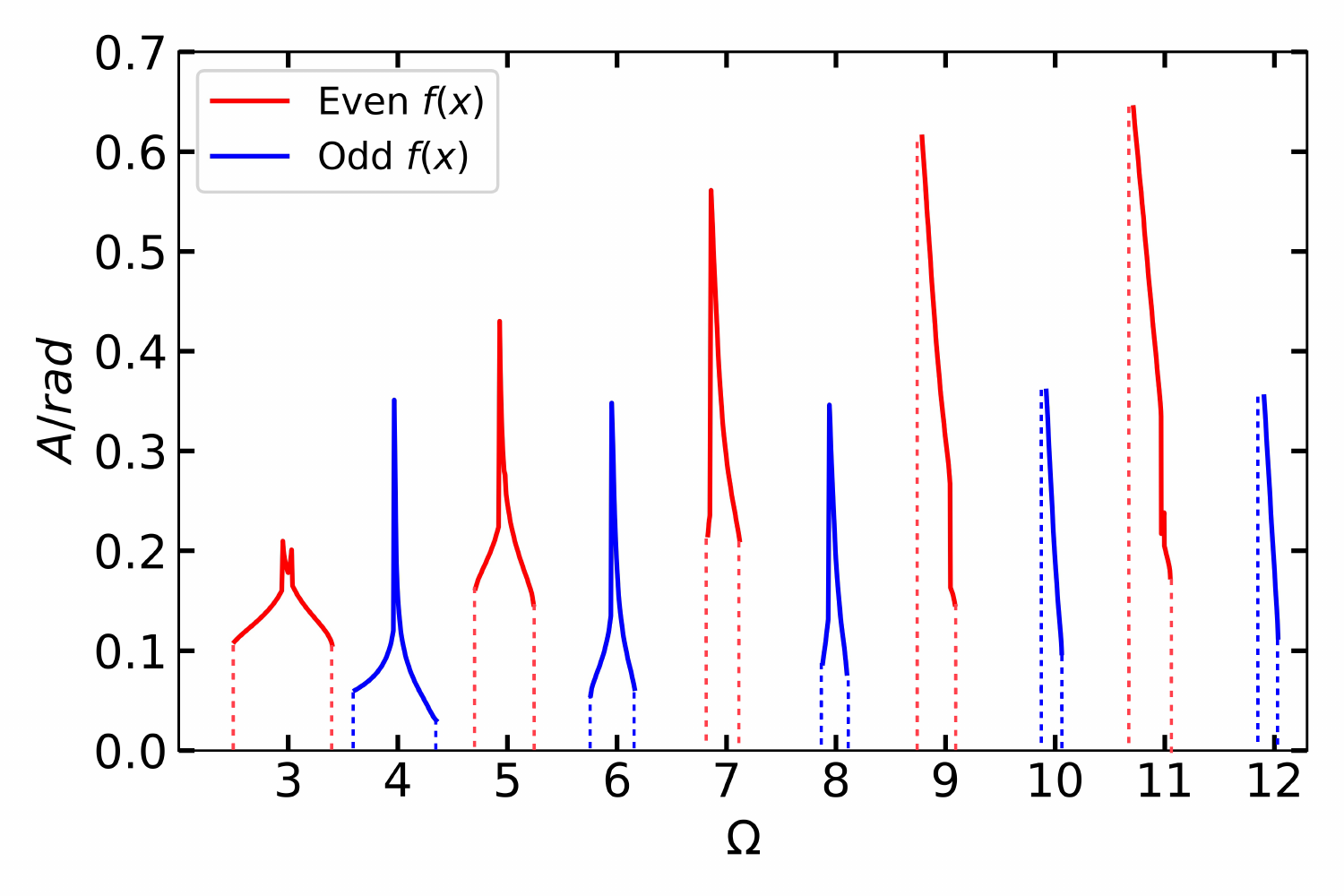}
  \caption{The max-amplitude frequency response diagram. The red (blue) curves are for even (odd) feeding function. Subharmonic resonance occurs for $\Omega \approx 2k + 1 (2k) $ in response to even (odd) feeding function. The frequency response curve for linear forced oscillation is not plotted.}
  \label{Fig_6}
\end{figure}
\subsection{Symmetry of the Driving Force and Frequency Response}\label{Frequency}

To clarify how symmetry of the feeding function influences the subharmonic frequencies, we calculate the overall frequency response diagram for even symmetric and odd symmetric $f(x)$. In calculation, we consider damping and using Eq.\ref{eq_8Driving force} and parameters listed in table.\ref{table} to model the feeding function, the frequency response diagram can be solved numerically. Since multiple periodic oscillation modes coexist when subharmonic resonance occurs, we only show maximum amplitudes on the frequency response diagram.

In Fig.\ref{Fig_6}, red curves correspond to even symmetric feeding function, and blue curves for odd symmetric feeding function. We find that subharmonic resonance occurs for $\Omega \approx 2k+1$ ($2k$) in response to even (odd) symmetric feeding function. Since an asymmetric function can be decomposed into an odd function and even function, we can infer that if the electromagnet is placed inclined, subharmonic resonance could occur for $\Omega \approx k$.



\begin{figure*}[t]
  \centering
     \subfigure[$\Omega =2.940$]{
    \label{7-0} 
    \includegraphics[width=0.21\textwidth]{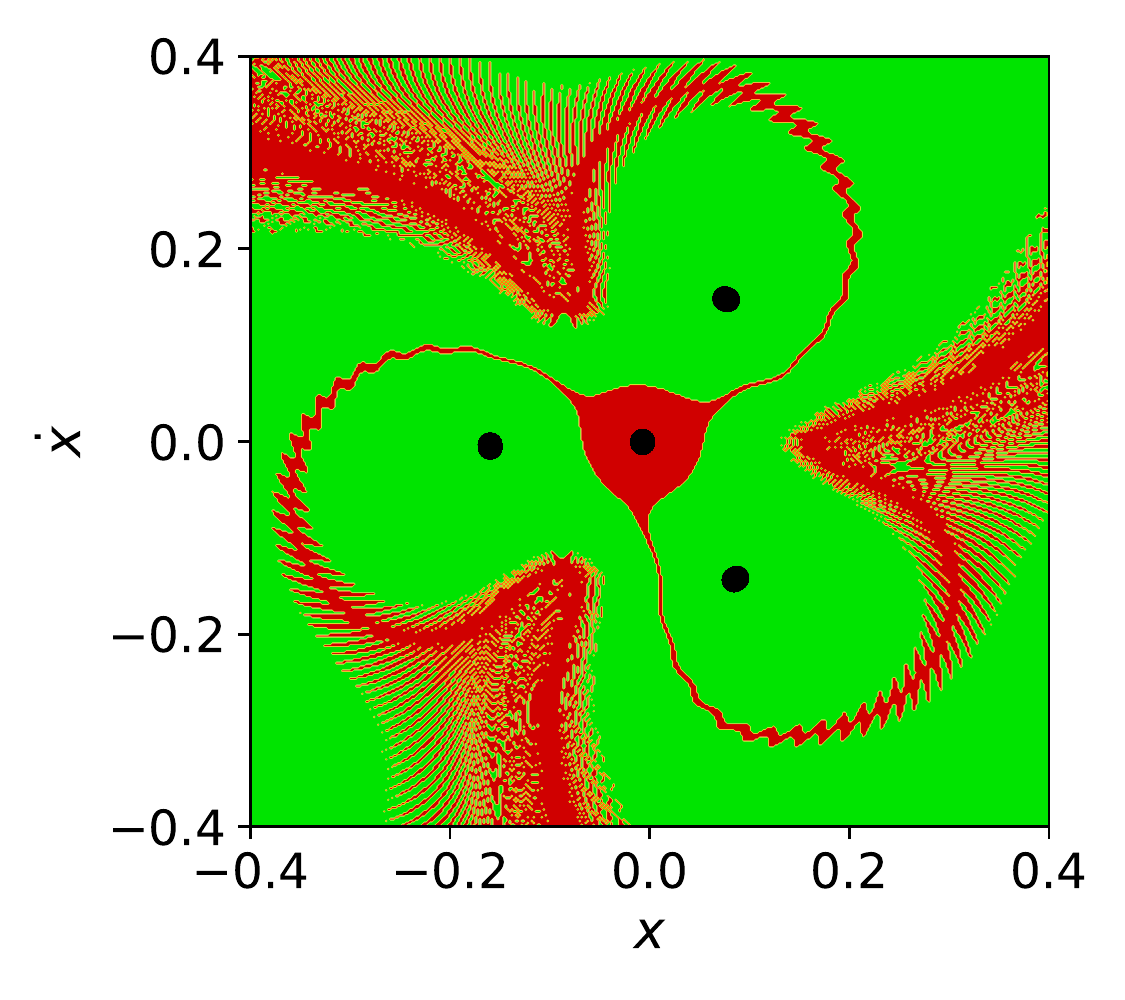}}
   \subfigure[$\Omega =2.945$]{
    \label{7-1} 
    \includegraphics[width=0.21\textwidth]{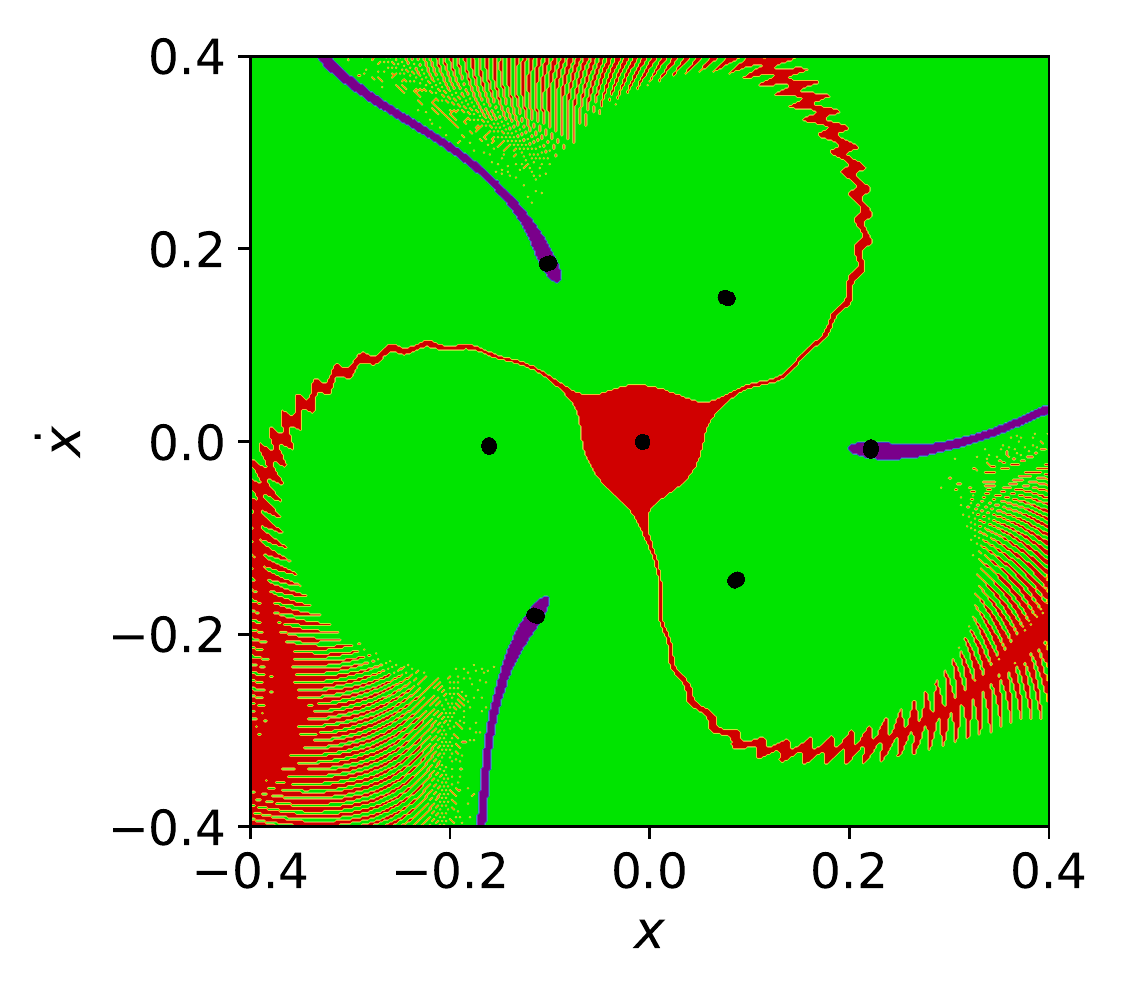}}
  \subfigure[$\Omega =3.000$]{
    \label{7-2} 
    \includegraphics[width=0.21\textwidth]{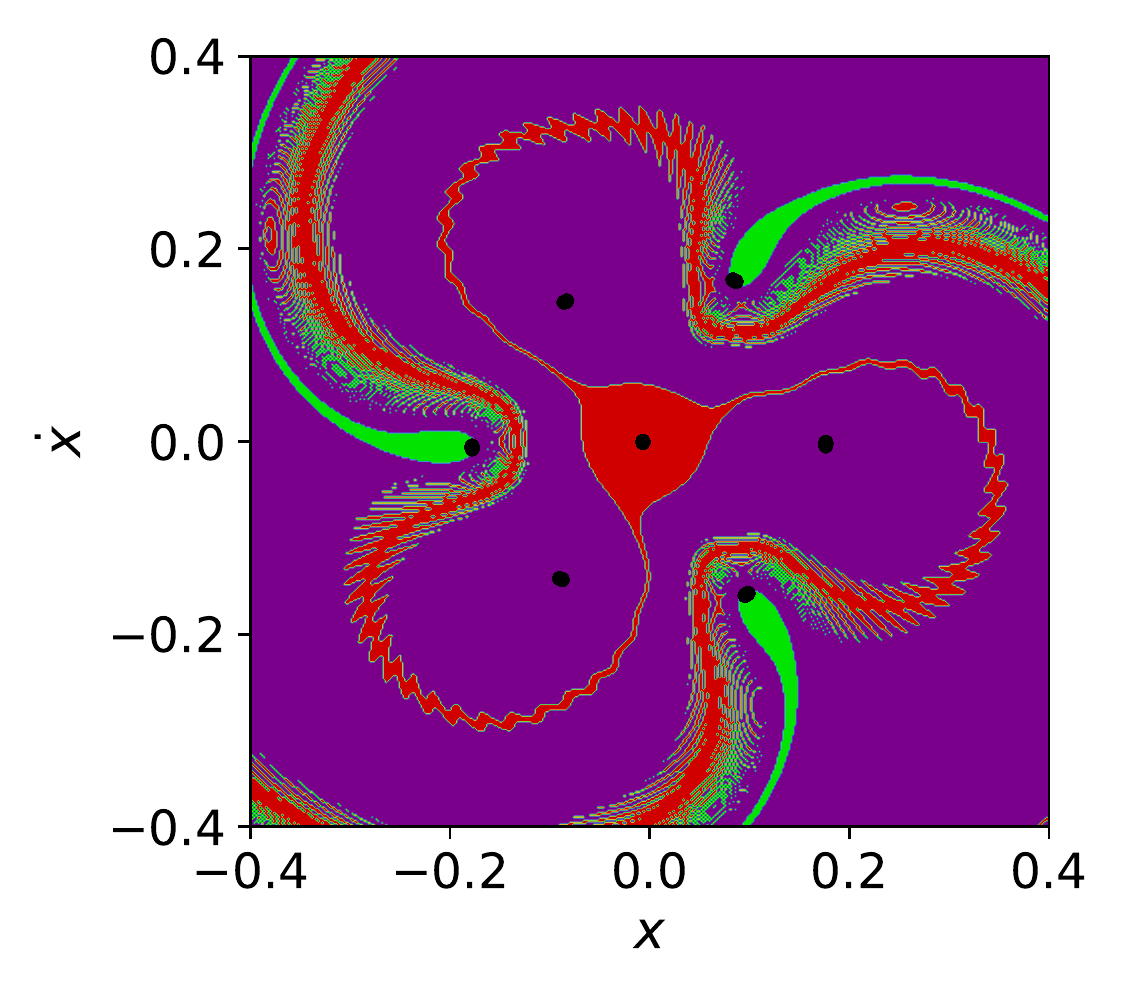}}
  \subfigure[$\Omega =3.040$]{
    \label{7-25} 
    \includegraphics[width=0.3\textwidth]{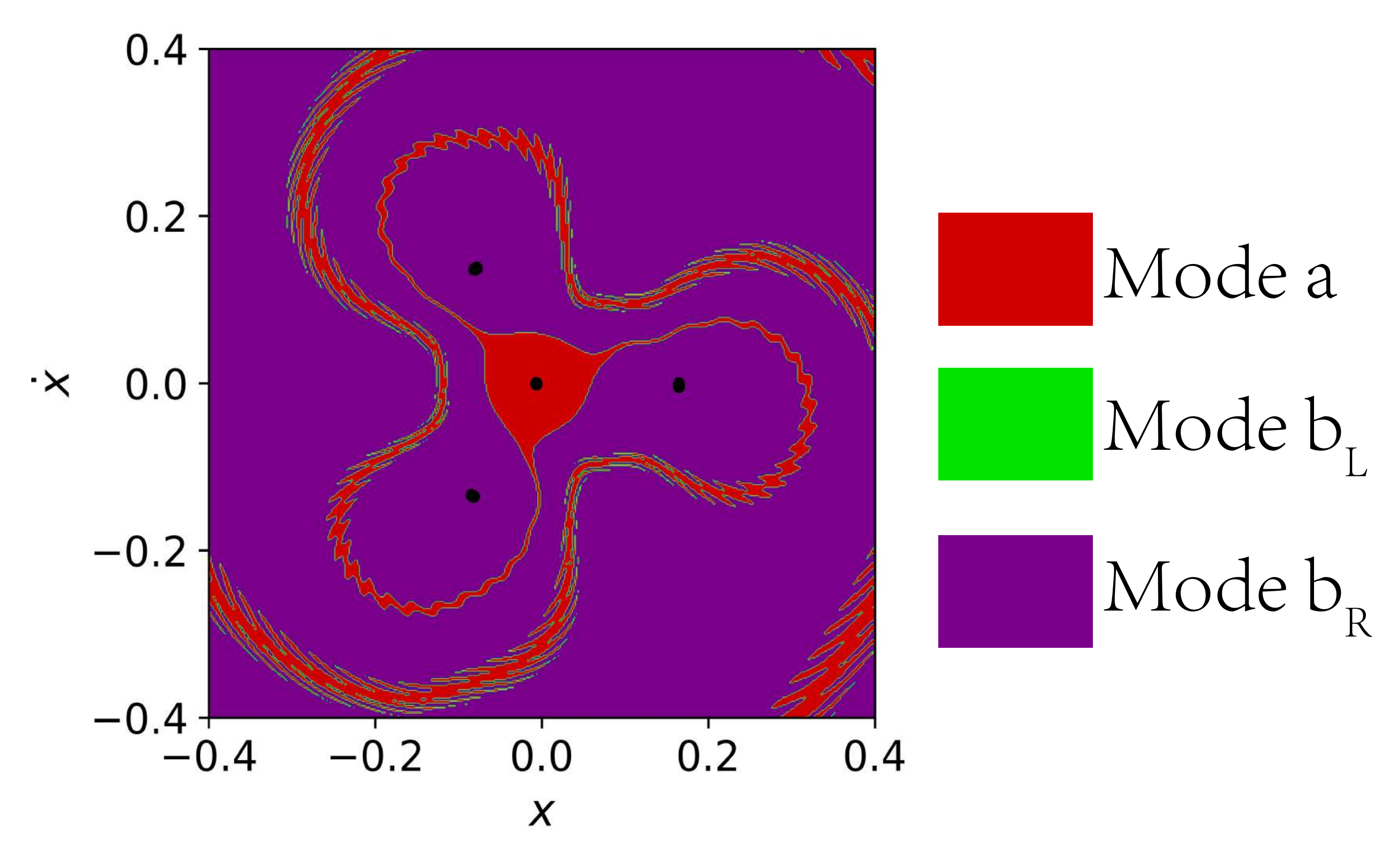}}
   \subfigure [$\Omega \approx 3$]{
    \label{7-3} 
    \includegraphics[width=0.3\textwidth]{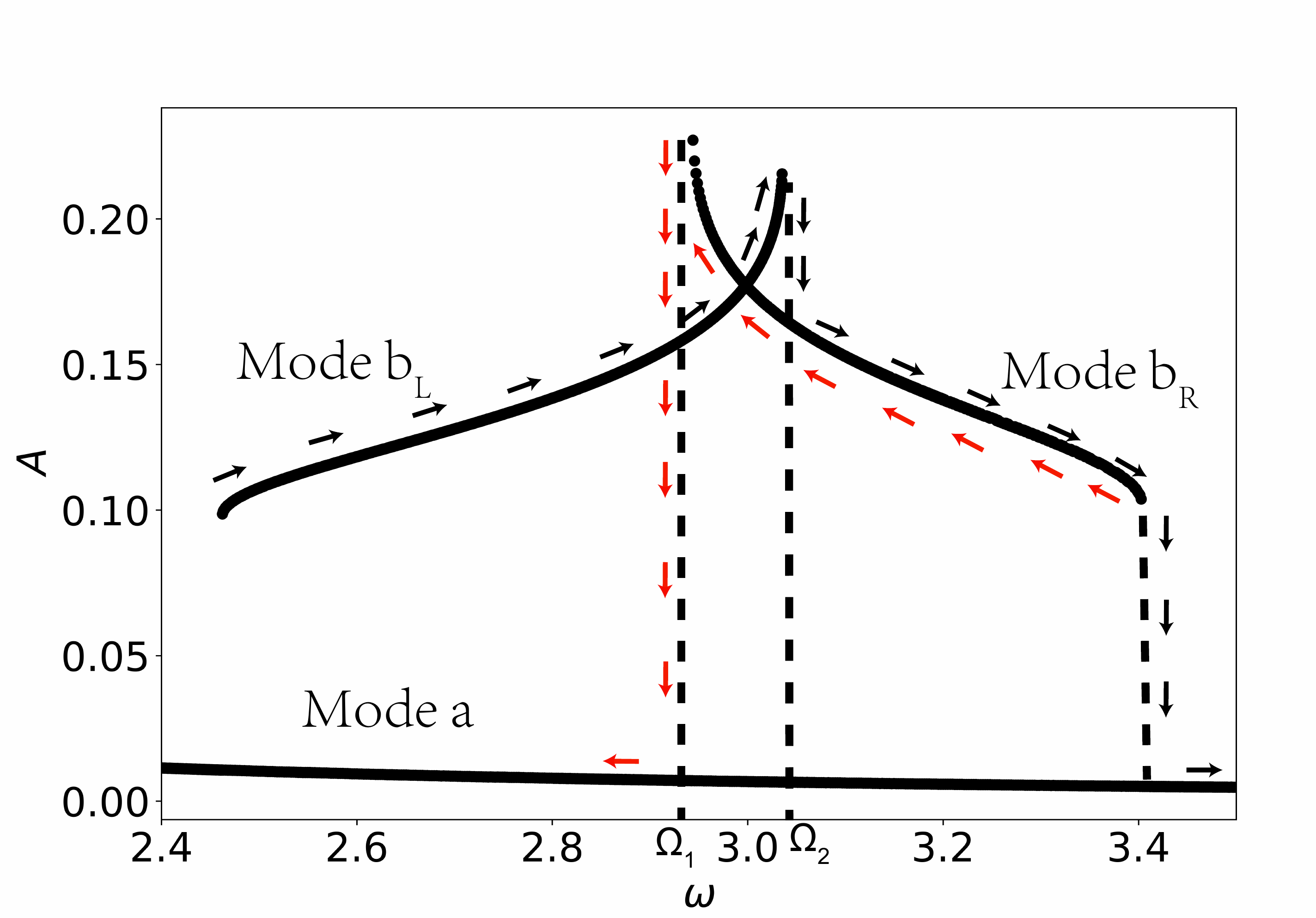}}

  \caption{\textbf{a},\textbf{b},\textbf{c},\textbf{d}, Evolution of attraction basins in Poincare maps with varying $\Omega$. The black dots are FPs for each basin. Attraction basin for each FP is denoted with a color. \textbf{e}, Frequency response curves for $\Omega \approx 3$. The arrows beside the frequency response curves indicate how oscillation modes evolve with driving frequency.}
  \label{Fig_7}
\end{figure*}

Fig.\ref{7-3} shows the frequency response diagram when $\Omega \approx 3$. Similar to analytical result in Fig.\ref{Fig_5}, it has a small linear solution, denoted as mode $a$, and a pair of intersecting branches denoted as mode $b_R$ and mode $b_L$ for $\Omega \approx 3$. $L$ and $R$ stand for right and left; $a$, $b$ and $c$ denote different branches. Here the unstable solutions are not presented because they are unavailable via numerical integral of Eq.\ref{eq6}.
Fig.\ref{8-4} is the frequency response diagram for $\Omega \approx 5$.
The five stable orbits in Fig.\ref{4-2}  correspond to the one linear branch named $a$ and two pairs of overlapped branches, named as $b_L,b_R,c_L,c_R $.

\subsection{Modes Competition}\label{modes}

A Duffing oscillator with cubic nonlinearity in restoring force\cite{kovacic2011duffing} exhibits jump phenomena and hysteresis between bistable states due to hardening or softening resonance. Meanwhile, the frequency response curves of a nonlinear parametric oscillator with bistability exhibit mixed feature of softening and hardening\cite{rhoads2006generalized}. Here, more complex jump phenomenon, induced by modes competition among multiple attractors, is studied.

In order to study the transition behavior among the attractors, the frequency response graph is calculated when excitation frequency increases or decreases quasi-statically.
Transition happens when one orbit loses stability and the oscillator jumps to another orbit, as the control parameter, i.e. the driving frequency, changing quasi-statically.
In Fig.\ref{7-3}, the arrows beside the frequency response curves indicate how transition happens.
 The frequency condition when all stationary modes $b_R$, $b_L$ and $a$ coexist is $\Omega_1 \leq \Omega \leq \Omega_2$. When $\Omega \leq \Omega_1$, modes $b_L$ and $a$ are stable, and when $\Omega > \Omega_2$, modes $b_R$ and $a$ are stable.  As frequency increases, the oscillator in mode $b_L$ jumps to mode $b_R$ at $\Omega=\Omega_2$. Unexpectedly, as frequency decreases, transition from mode $b_R$ to mode $a$ happens at $\Omega=\Omega_1$.

Such irreversible transition behavior can be described from the evolution of attraction basins in Poincare maps.
Figs.\ref{7-0} to \ref{7-25} are the attraction basins when $\Omega= 2.94, 2.945, 3.0, 3.04$ respectively. In each figure, the small black dots are fixed points (FPs). Each FP is surrounded by its basin of attraction indicated with color. The red, green and purple areas denote the attraction basins for modes $a$, $b_L$ and $b_R$ respectively. Mode $a$ (the trivial solution) leaves one fixed point near the center in Figs.\ref{7-0} to \ref{7-25}. Meanwhile, each large-amplitude mode has 3 fixed points because the oscillation periods of modes $b_R$ and $b_L$ are triple of excitation period.




In Figs.\ref{7-1}, \ref{7-2} and \ref{7-25}, as the excitation frequency increases from $\Omega=2.945$ to 3.00 and 3.040, the attraction basin of $b_L$ shrinks quickly and vanished below 3.040. In Fig.\ref{7-2} just below the transition frequency, the vanishing attraction basins of $b_L$ surrounded by those of $b_R$ will merged into those of $b_R$ and transition from $b_L$ to $b_R$ happens.

In Figs.\ref{7-1} and \ref{7-0}, in vicinity of the critical frequency $\Omega_1$, the basins of $b_R$ are surrounded by those of mode $a$ (the red area). As the excitation frequency decreases from 2.945 to 2.940, the basins of mode $b_R$ are replaced by those of mode $a$ so that transition from mode $b_R$ to $a$ happens.




\begin{figure*}[htpb]
  \centering
     \subfigure[$\Omega =4.92$]{
    \label{8-0} 
    \includegraphics[width=0.21\textwidth]{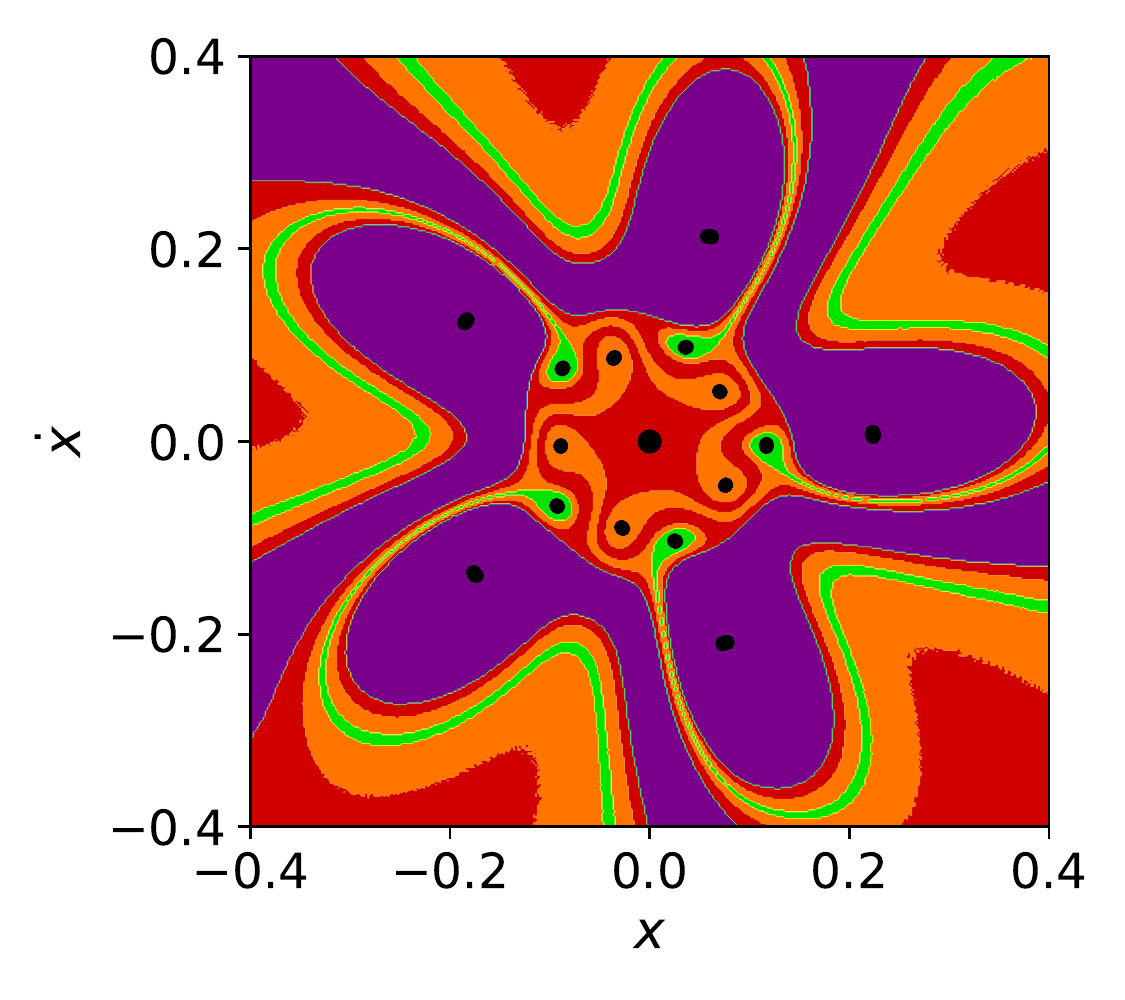}}
   \subfigure[$\Omega =4.95$]{
    \label{8-1} 
    \includegraphics[width=0.21\textwidth]{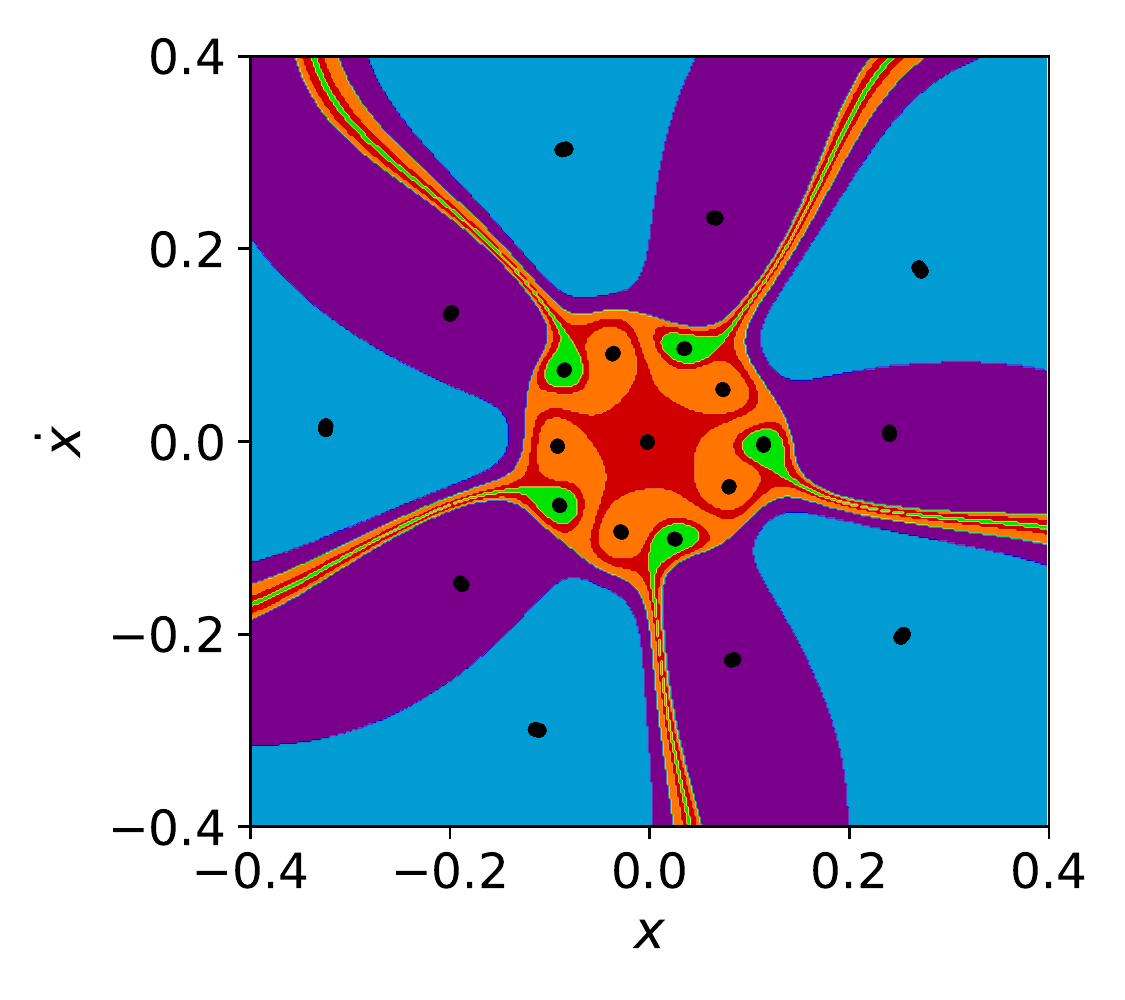}}
  \subfigure[$\Omega =4.98$]{
    \label{8-2} 
    \includegraphics[width=0.21\textwidth]{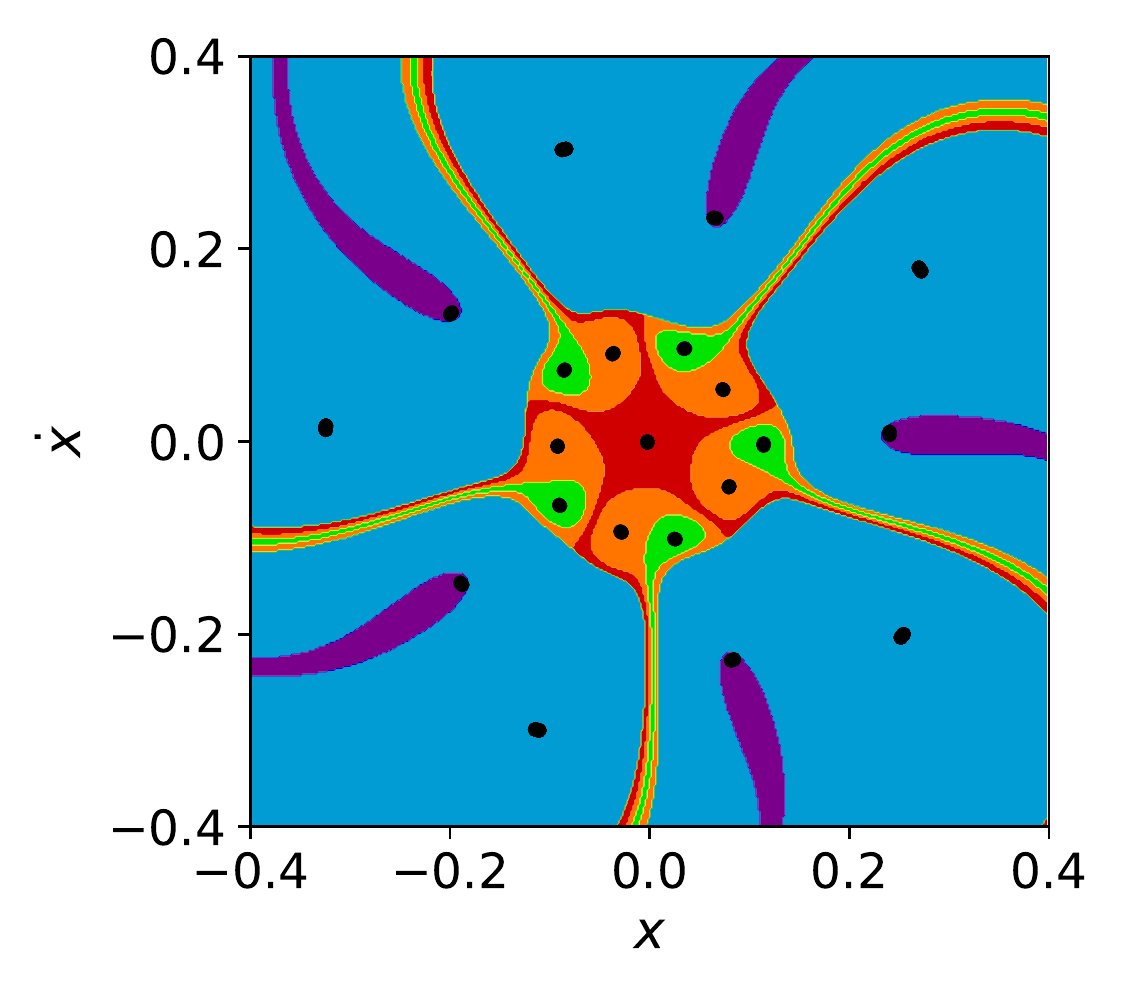}}
   \subfigure[$\Omega =5.00$]{
    \label{8-3} 
    \includegraphics[width=0.3\textwidth]{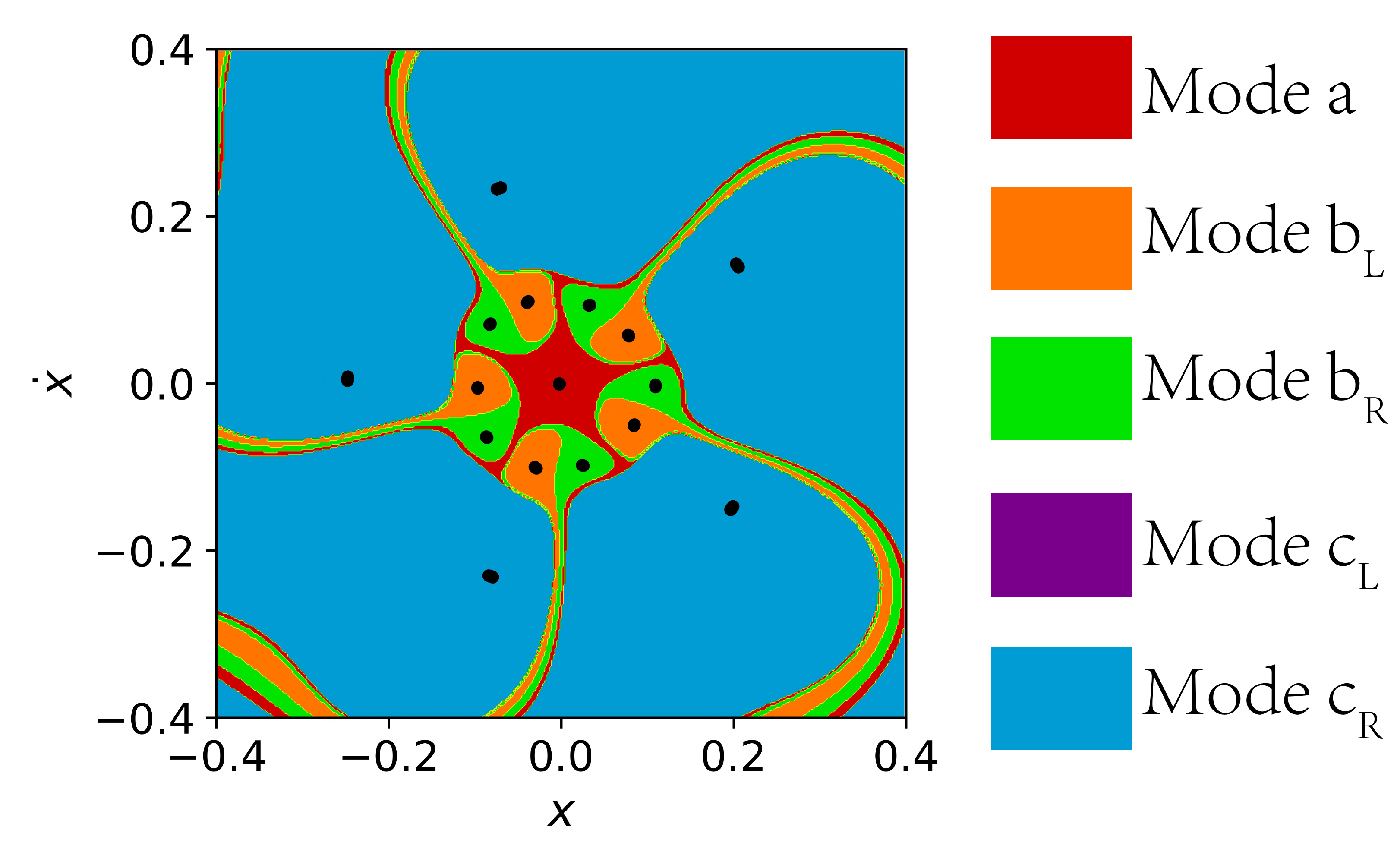}}
   \subfigure [$\Omega \approx 5$]{
    \label{8-4} 
    \includegraphics[width=0.3\textwidth]{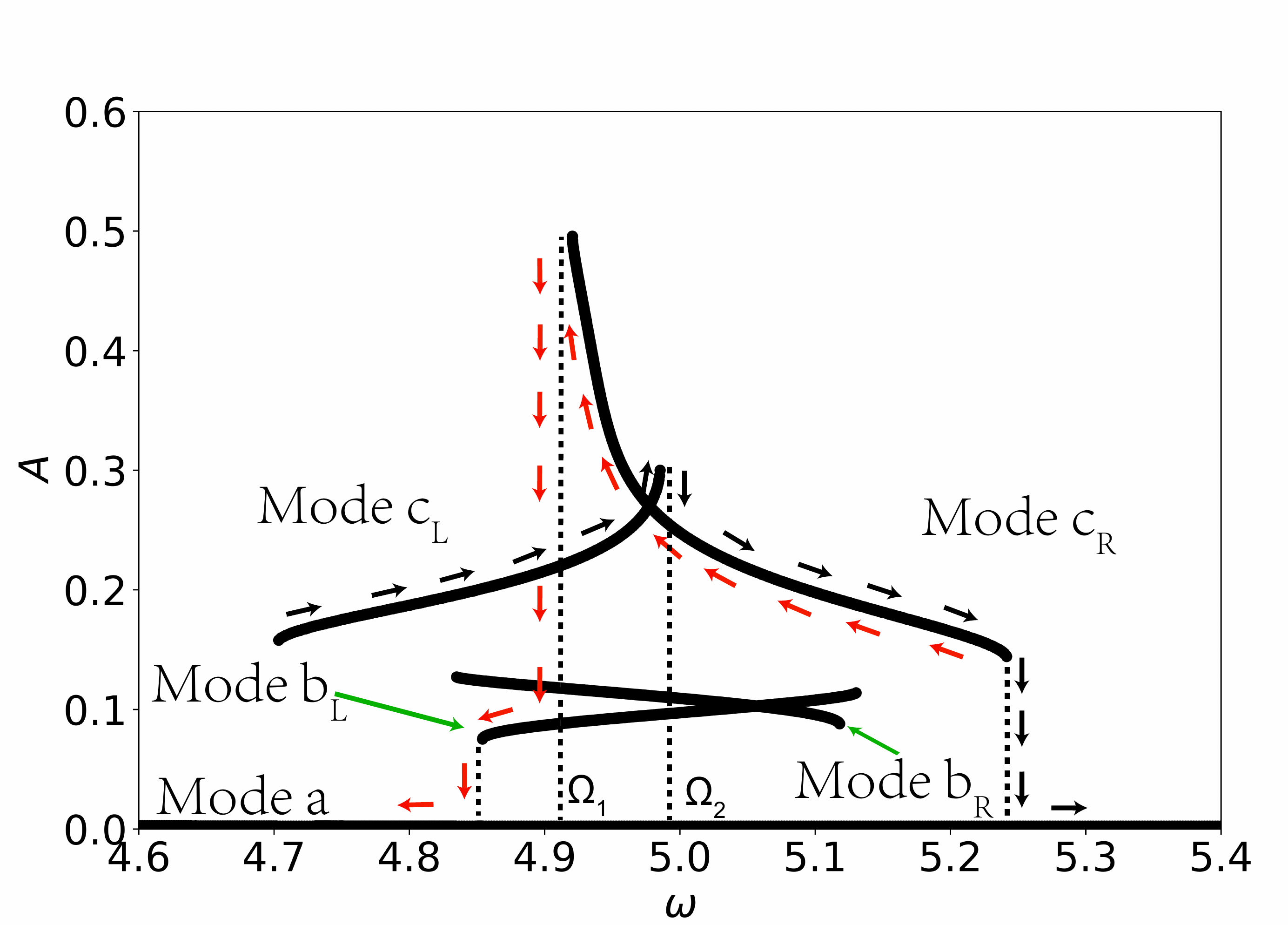}}
  \caption{ \textbf{a},\textbf{b},\textbf{c},\textbf{d},
  Evolution of the attraction basins in Poincare maps for $\Omega = 4.92,4.95,4.98,5.0$. The black dots are FPs for each basin. Attraction basin for each FP is denoted with a color.\textbf{e}, Frequency response curves for $\Omega \approx 5$. The arrows beside the frequency response curves indicate how oscillation modes evolve with driving frequency.}
  \label{Fig_8}
\end{figure*}

 Additionally, the frequency response diagram and Poincare maps for $\Omega \approx 5$ are presented, see Fig.\ref{Fig_8}.
$\Omega_1$ is the lower critical frequency for mode $c_R$, and $\Omega_2$ is the upper critical frequency for mode $c_L$. In Figs.\ref{8-0} to \ref{8-3}, the red, orange, green, purple and blue areas are the attraction basins of modes $a$, $b_L$, $b_R$, $c_L$ and $c_R$ respectively. We mainly focus on a typical transition starting from model $c_L$ and mode $c_R$ below.

Fig.\ref{8-4} shows that transition among different modes is also irreversible. When increasing the driving frequency, oscillator at mode $c_L$ will jump to mode $c_R$ at $\Omega=\Omega_2$. In Fig.\ref{8-2}, the attraction basins of mode $c_L$ are surrounded by those of $c_R$ below $\Omega_2$, and the basins of $c_L$ will merge into the $c_R$ basin as driving frequency exceeds $\Omega_2$, see Fig.\ref{8-3}.


When decreasing the driving frequency, oscillator at mode $c_R$ jumps to mode $b_L$ at $\Omega=\Omega_1$, rather than $c_L$. In Fig.\ref{8-1} and Fig.\ref{8-0}, the attraction basins of mode $c_R$ will merge into the basin of mode $b_L$ below $\Omega_1$.


From the transition sequence described in Fig.\ref{7-3} and Fig.\ref{8-4}, we propose a control strategy among the oscillation modes by adiabatically changing the driving frequency. For example, in Fig.\ref{7-3} starting from mode $b_L$, the oscillator can be induced to mode $b_R$ by increasing the driving frequency, then to mode $a$ by decreasing driving frequency. However, the transition from mode $a$ to higher oscillation modes are forbidden. It means that an oscillator initially performs linear oscillation in mode $a$, cannot be driven to higher orbits merely by changing the driving frequency.

\section{Conclusion}
The phenomena and mechanisms of multistability involve rich dynamics to be explored. In this paper, we investigate the multistability of a generalized nonlinear forcing oscillator excited by $f(x)cos \Omega t$. We take Doubochinski's Pendulum as an example. The multistability mechanism in our system is nonlinear parametric resonance. We express the energy feeding function into polynomial and find multi-stability can be achieved without dissipation when the polynomial contains more than one nonlinear terms so that energy balance can be realized. We also find the subharmonic resonance frequency conditions and demonstrate the even-odd correspondence between the symmetry of the feeding function and subharmonic resonance frequencies.


We present the frequency response diagrams and Poincare maps near subharmonic frequencies $3 \omega_0$ and  $5 \omega_0$. The frequency response diagram contains one linear response branch and nonlinear branches appearing in pairs. We find irreversible transition phenomenon between the multistable modes.
Control of the multi-stability can be realized by changing the excitation frequency adiabatically.


\begin{thebibliography}{24}
\providecommand{\natexlab}[1]{#1}
\providecommand{\url}[1]{\texttt{#1}}
\expandafter\ifx\csname urlstyle\endcsname\relax
  \providecommand{\doi}[1]{doi: #1}\else
  \providecommand{\doi}{doi: \begingroup \urlstyle{rm}\Url}\fi

\bibitem{maurer1980effect}
J.~Maurer and A.~Libchaber, Effect of the prandtl number on the onset of
  turbulence in liquid 4he, \emph{Journal de Physique lettres}. {\bf
  41}\penalty0 (21), \penalty0 515--518  (1980).

\bibitem{brun1985observation}
E.~Brun, B.~Derighetti, D.~Meier, R.~Holzner, and M.~Ravani, Observation of
  order and chaos in a nuclear spin--flip laser, \emph{JOSA B}. {\bf
  2}\penalty0 (1), \penalty0 156--167  (1985).

\bibitem{gibbs2012optical}
H.~Gibbs, \emph{Optical bistability: controlling light with light}. Elsevier
  (2012).

\bibitem{aguda1987bistability}
B.~D. Aguda and B.~L. Clarke, Bistability in chemical reaction networks: theory
  and application to the peroxidase--oxidase reaction, \emph{The Journal of
  chemical physics}. {\bf 87}\penalty0 (6), \penalty0 3461--3470  (1987).

\bibitem{wilhelm2009smallest}
T.~Wilhelm, The smallest chemical reaction system with bistability, \emph{BMC
  systems biology}. {\bf 3}\penalty0 (1), \penalty0 90  (2009).

\bibitem{angeli2004detection}
D.~Angeli, J.~E. Ferrell, and E.~D. Sontag, Detection of multistability,
  bifurcations, and hysteresis in a large class of biological positive-feedback
  systems, \emph{Proceedings of the National Academy of Sciences}. {\bf
  101}\penalty0 (7), \penalty0 1822--1827  (2004).

\bibitem{ozbudak2004multistability}
E.~M. Ozbudak, M.~Thattai, H.~N. Lim, B.~I. Shraiman, and A.~Van~Oudenaarden,
  Multistability in the lactose utilization network of escherichia coli,
  \emph{Nature}. {\bf 427}\penalty0 (6976), \penalty0 737  (2004).

\bibitem{freyer2011biophysical}
F.~Freyer, J.~A. Roberts, R.~Becker, P.~A. Robinson, P.~Ritter, and
  M.~Breakspear, Biophysical mechanisms of multistability in resting-state
  cortical rhythms, \emph{Journal of Neuroscience}. {\bf 31}\penalty0 (17),
  \penalty0 6353--6361  (2011).

\bibitem{robinson2012multistability}
A.~Robinson, R.~Calov, and A.~Ganopolski, Multistability and critical
  thresholds of the greenland ice sheet, \emph{Nature Climate Change}. {\bf
  2}\penalty0 (6), \penalty0 429  (2012).

\bibitem{freire2008multistability}
J.~G. Freire, C.~Bonatto, C.~C. DaCamara, and J.~A. Gallas, Multistability,
  phase diagrams, and intransitivity in the lorenz-84 low-order atmospheric
  circulation model, \emph{Chaos: An Interdisciplinary Journal of Nonlinear
  Science}. {\bf 18}\penalty0 (3), \penalty0 033121  (2008).

\bibitem{pisarchik2014control}
A.~N. Pisarchik and U.~Feudel, Control of multistability, \emph{Physics
  Reports}. {\bf 540}\penalty0 (4), \penalty0 167--218  (2014).

\bibitem{arecchi1982experimental}
F.~Arecchi, R.~Meucci, G.~Puccioni, and J.~Tredicce, Experimental evidence of
  subharmonic bifurcations, multistability, and turbulence in a q-switched gas
  laser, \emph{Physical Review Letters}. {\bf 49}\penalty0 (17), \penalty0 1217
   (1982).

\bibitem{maistrenko2007multistability}
Y.~L. Maistrenko, B.~Lysyansky, C.~Hauptmann, O.~Burylko, and P.~A. Tass,
  Multistability in the kuramoto model with synaptic plasticity, \emph{Physical
  Review E}. {\bf 75}\penalty0 (6), \penalty0 066207  (2007).

\bibitem{balanov2005delayed}
A.~G. Balanov, N.~B. Janson, and E.~Sch{\"o}ll, Delayed feedback control of
  chaos: Bifurcation analysis, \emph{Physical Review E}. {\bf 71}\penalty0 (1),
  \penalty0 016222  (2005).

\bibitem{foss1996multistability}
J.~Foss, A.~Longtin, B.~Mensour, and J.~Milton, Multistability and delayed
  recurrent loops, \emph{Physical Review Letters}. {\bf 76}\penalty0 (4),
  \penalty0 708  (1996).

\bibitem{varma1993quadratic}
V.~Varma et~al., Quadratic map modulated by additive periodic forcing,
  \emph{Physical Review E}. {\bf 48}\penalty0 (3), \penalty0 1670  (1993).

\bibitem{lieberman1985transient}
M.~A. Lieberman and K.~Y. Tsang, Transient chaos in dissipatively perturbed,
  near-integrable hamiltonian systems, \emph{Physical review letters}. {\bf
  55}\penalty0 (9), \penalty0 908  (1985).

\bibitem{feudel1996map}
U.~Feudel, C.~Grebogi, B.~R. Hunt, and J.~A. Yorke, Map with more than 100
  coexisting low-period periodic attractors, \emph{Physical Review E}. {\bf
  54}\penalty0 (1), \penalty0 71  (1996).

\bibitem{tennenbaum2006amplitude}
J.~Tennenbaum, Amplitude quantization as an elementary property of macroscopic
  vibrating systems, \emph{21ST CENTURY SCIENCE AND TECHNOLOGY}. {\bf
  18}\penalty0 (4), \penalty0 50  (2006).

\bibitem{nayfeh2008nonlinear}
A.~H. Nayfeh and D.~T. Mook, \emph{Nonlinear oscillations}. John Wiley \& Sons
  (2008).

\bibitem{kovacic2011duffing}
I.~Kovacic and M.~J. Brennan, \emph{The Duffing equation: nonlinear oscillators
  and their behaviour}. John Wiley \& Sons  (2011).

\bibitem{rhoads2006generalized}
J.~F. Rhoads, S.~W. Shaw, K.~L. Turner, J.~Moehlis, B.~E. DeMartini, and
  W.~Zhang, Generalized parametric resonance in electrostatically actuated
  microelectromechanical oscillators, \emph{Journal of Sound and Vibration}.
  {\bf 296}\penalty0 (4-5), \penalty0 797--829  (2006).

\bibitem{damgov2000discrete}
V.~Damgov and I.~Popov, “discrete” oscillations and multiple attractors in
  kick-excited systems, \emph{Discrete Dynamics in Nature and Society}. {\bf
  4}\penalty0 (2), \penalty0 99--124  (2000).

\bibitem{njubook}
W.~G. Sihui~Wang, \emph{International Young Physicists' Tournament: Problems
  and Solutions 2015}. WorldScientific  (2018).

\end{thebibliography}
\end{document}